# Brake wear (nano)particle characterization and toxicity on airway epithelial cells *in vitro*

**Chloé Puisney[1,2], Evdokia K. Oikonomou[2], Sophie Nowak[3], Alexandre Chevillot[3], Sandra Casale[4], Armelle Baeza-Squiban[1*] and Jean-François Berret[2*]**

[1]*Université Paris Diderot, Université Sorbonne Paris Cité, UMR CNRS 8251, Laboratoire de Réponses Moléculaires et Cellulaires aux Xénobiotiques, Paris, France*
[2]*Université Paris Diderot, Université Sorbonne Paris Cité, Laboratoire Matière et Systèmes Complexes, UMR 7057 CNRS, Paris, France*
[3]*Université Paris Diderot, Université Sorbonne Paris Cité, UMR CNRS 7086, Laboratoire Interfaces, Traitements, Organisation et Dynamique des Systèmes, Paris, France*
[4]*Université Pierre et Marie Curie, Institut des Matériaux de Paris Centre, Paris, France*

**Keywords**
Air pollution, Brake wear particles, Dynamic light scattering, Inflammation, Metallic particle, Respiratory epithelium

**Corresponding authors**
Armelle BAEZA-SQUIBAN: Univ Paris Diderot, Sorbonne Paris Cité, Unit of Functional and Adaptive Biology, UMR 8251 CNRS, 5 rue Thomas Mann, F-75205, Paris, France.
baeza@univ-paris-diderot.fr

Jean-François BERRET, Univ Paris Diderot, Sorbonne Paris Cité, Laboratoire Matière et Systèmes Complexes, UMR 7057 CNRS, 10 rue Alice Domon et Léonie Duquet, 75205 Paris, jean-francois.berret@univ-paris-diderot.fr

**Abstract** Particulate air pollution results from different sources, among which those related to road traffic have a significant impact on human health. Combustion-derived particles emitted by thermal engines have been incriminated and are now better controlled. In contrast, non-exhaust emission sources related to car wear and degradation processes are not yet regulated. Here we report on brake wear particles (BWP) harvested in two test facilities operating in France, providing samples from different braking systems and driving/testing conditions. Using a combination of light scattering, X-ray fluorescence, optical and electron microscopy, the particle size and elemental composition are revealed. The BWP are shown to be in the nano- to micrometer range and to have a low carbonaceous content (6%), iron and copper being the main components (> 40%). To evaluate the toxicity potential of its nano-sized fraction, brake wear nanoparticles are isolated by sonication, filtration and ultra-centrifugation techniques, leading to stable colloidal dispersions. A significant outcome of this study is that the nano-sized fraction represents 26% by mass of the initial BWP. Human bronchial epithelial cells (Calu-3) are used as relevant target cells to investigate their cytotoxicity. We observe a clear short-term loss of viability associated to reactive oxygen species generation, but with limited pro-inflammatory effects. On an actual cell-deposited mass-dose basis, the cytotoxicity of the nanosized fraction is similar to that of BWP, suggesting that the cytotoxicity is particle size independent. To conclude, brake wear dust contains substantial amount of metallic nanoparticles exhibiting toxicity for lung cells, and should warrant further consideration.

**Environmental significance**

Air pollution is now recognized as a major environment-related health threat. Traffic exhaust particles are considered to contribute significantly to this threat, resulting in the implementation of specific regulations for the automotive industry. Until now little attention has been paid to non-exhaust particles, especially to wear particles emitted by transport vehicles during braking. Here we provide evidences that brake wear particles coming from two distinct sources contain large amounts of metallic particles of size less than 100 nm. These nanoparticles are shown to be cytotoxic towards human bronchial cells, but display limited pro-inflammatory effects. The presence of metallic nanoparticles in brake wear emission raises concerns about their ability to translocate through the respiratory barrier and trigger systemic effects.

# Introduction

Air pollution is now recognized as a major environment-related health threat. A number of epidemiological studies have shown the existence of an association between the exposure to air pollution and the increase of mortality and morbidity from cardiorespiratory diseases.[1,2] Toxicological studies have supported the biological plausibility of this association, especially for the particulate matter (PM) component.[3] PM air pollution results from different sources, among which road traffic has a significant part, impacting urban populations worldwide. Combustion-derived particles emitted by thermal engines are specifically held accountable, which has led state agencies to implement stricter regulations for the automotive industry, eventually resulting in the reduction of the $PM_{2.5}$ (particles with aerodynamic diameters ≤ 2.5 μm) in the last decades.[4] By contrast, non-exhaust emission sources related to wear mechanisms are not yet regulated. Wear and degradation processes concern the subcomponent moving parts in cars, including the clutch, tires, brakes etc. as well as the direct driving environments, such as roads or tarmac. Moreover, with the broadening of exhaust post-treatment (e.g. diesel particulate filters) and the emergence of electric vehicles, the proportion of wear particles in urban pollution is expected to soar in the coming years.[5,6]

Recent studies have shown that the contribution of brake wear particles (BWP) can represent up to 55% by mass of the total non-exhaust traffic-related $PM_{10}$ emission and up to 21% by mass of the total traffic related $PM_{10}$ (particles with aerodynamic diameters ≤ 10 μm) emissions.[7,8] The two types of frictional braking systems available for light-duty and small vehicles are the drum and disc/pad brakes, whose emission have been quantified in specific contexts at a level of 30 mg $km^{-1}$ and 9 mg $km^{-1}$ of particles per vehicle, respectively.[9] There exists many different brands of braking systems on the market, and most of them comprise similar lining components, such as binders, fillers, reinforcing agents (fibers) and friction modifiers (abrasives and lubricants).[10] The filler and fiber components as well as some additives are mainly metallic, whereas the binders and additives are carbonaceous.[10,11] During braking, the friction between the pad and disc surfaces is used to convert the vehicle kinetic energy into heat, resulting in a significant alteration of the lining components and in a concomitant temperature increase.[8] The temperature there can grow to 400 °C and locally it may reach up to 700 to 1000 °C, depending on the testing or driving conditions.[12-14] In this process, metallic particles in the nano- to micrometer range are generated, together with carbon-based particles stemming from the condensation of evaporated organic compounds.[9] BWPs may have an impact on the environment and human health due to their composition and size. In particular, the nanoscale fraction of the BWPs could be preferentially retained in the vicinity of a roadway compared to larger more readily settleable materials. As such, the inhaled dose and exposure could be much higher than for larger particles. Traffic-related wear emissions are nowadays considered relevant for air quality policy development, however for now no specific regulations have been implemented.

More generally pertaining to human health, BWP emission leads to airborne (nano)particles for which the preferential entrance pathway is the respiratory tract. Besides their local impact on the lung epithelia, nanoparticles reaching the alveolar region may cross the air-blood barrier and accumulate in secondary organs.[15,16] So far, few studies have dealt explicitly with the toxicity of BWPs, and most of them were limited to *in vitro* assays, and not focused on the nano-sized fraction. In general it was found that their high content in transition metals and metallic compounds can generate various cytotoxic effects.[17,18] With the A549 alveolar cell line, BWPs were shown to induce oxidative stress[19] and to alter the tight junctions between the cells.[18] On peripheral blood lymphocytes, brake wear dust was found to cause chromosomal damages, likely produced by the metallic nanocrystals present in the dust.[20]

Our present work focuses on pooled BWP samples collected on two test center facilities in France and corresponding to different braking conditions. The objectives are to characterize the BWPs in terms of size and composition, and to isolate its nano-sized fraction by a combination of sonication, filtration and ultra-centrifugation techniques.[21] The resulting nanoparticle dispersions in water are found to be stable over time and they are tested on a model of human bronchial epithelial cells (Calu-3). As a result, a short-term loss of viability associated to reactive oxygen species generation is observed. The nano-sized fraction is found to exhibit lower toxicity and reactive oxygen species production than BWPs. In terms of effective dose however, the toxicity levels are indeed similar, suggesting that under the conditions of the study the cytotoxicity is particle size independent.

# Materials and Methods

## Brake wear particle collection

BWPs were obtained from two different sources, a motor vehicle testing center operating on light duty vehicles and a test bench working with an automotive brake dynamometer. In the latter case, the dynamometer is equipped with an environmental chamber, *i.e.* the brake system is operating in close mode. In both facilities, the brake wear dust was collected in a similar manner. The wheel was dismounted and the brake system opened. The dust was harvested in a 50 mL Falcon tube using a clean and sterile spatula. The disc and pad surfaces were not scrapped in the procedure. For the dust coming from transport vehicles, different types of driving conditions were realized, including tests with antilock braking and traction control system, park assist, wiper and accessories on different road surfaces. This multiple driving conditions allow retrieving a representative mixture of BWPs emitted by the current European car fleet and under real-life driving cycles. In both facilities, the dust was harvested on a regular basis and pooled with previous collected samples. As a result, it comprised both newly generated fresh particles and particles that aged in the braking systems and were hence subject to oxidation or decomposition.[19] The workers collecting the samples were trained to minimize contamination during the operation.

**Brake wear particle characterization**

*X-ray fluorescence analysis:* The BWP elemental analysis was carried out by X-ray fluorescence spectrometry. The as-received samples were deposited on a low absorption prolene membrane and analyzed on an Epsilon 3XL (Panalytical) spectrometer equipped with a silver X-ray tube. The analysis was performed using the Omnian software which detects and quantifies elements from sodium to uranium.

*Thermogravimetric analysis:* Thermogravimetry was performed with LABSYS evo thermogravimetric analyzer (Setaram Instrumentations, Caluire-et-Cuire, France) under helium atmosphere. Approximately 50 mg of the as-received BWPs were placed in an alumina crucible and the temperature was raised from 30 °C to 1500 °C at a rate of 5 °C per minute.

*Scanning electron microscopy:* Scanning electron microscopy was performed on BWPs without metallization using a Zeiss SUPRA 40V SEM-FEG microscope (Oberkochen, Germany). The microscope was operated using an in-lens secondary electron detector working with a 2 kV acceleration tension and a 2 – 3 mm working distance.

*Optical microscopy:* Phase-contrast images were acquired on an IX73 inverted microscope (Olympus Corporation, Tokyo, Japan) equipped with 20×, 40×, and 60× objectives. Seven microliters of a dispersion were deposited on a glass plate and sealed into a Gene Frame (ABgeneAdvanced Biotechnologies, Hamburg, Germany) dual adhesive system.[22] An EXi Blue camera (QImaging, Surrey, Canada) and Metaview software (Universal Imaging Inc., Bedford Hills, USA) were used as the acquisition system. ImageJ software was used for scale bar (http://rsbweb.nih.gov/ij/).

**Brake wear dispersion and characterization**

*Nanoparticle dispersion*: BWPs were suspended in water at 1 mg mL$^{-1}$, vortexed and sonicated for 1 to 10 minutes at maximum power (10.7 W) using a cup-horn equipped sonicator (Branson Ultrasonics, Shangai, China). For each assay, 7 vials each containing 2 mL of suspension were placed in the cup and sonicated over time. This procedure was optimized to establish a reliable protocol, in particular concerning the particle size, charge and long-term stability. In a second phase, the protocol was repeated to produce large volumes of dispersion required for the physico-chemical and toxicological studies. To separate the nano-sized fraction, the dispersions were hereafter filtered with a syringe equipped with 0.45 µm Nalgene™ polyether sulfone membranes (Thermo Fisher Scientific, Waltham, USA).

*Light scattering:* Scattering intensities $I_S$ and hydrodynamic diameters $D_H$ were measured using a Zetasizer NanoZS equipment (Malvern Instruments, Worcestershore, United Kingdom). A 4 mW He−Ne laser beam ($\lambda$ = 633 nm) was used to illuminate sample dispersion and the scattered intensity was collected at an angle of 173°. The second-order autocorrelation function was analyzed using the CONTIN algorithm to determine the average diffusion coefficient $D_C$ of the scatterers. The hydrodynamic diameter was calculated according to the Stokes−Einstein relation, $D_H = k_B T / 3\pi\eta D_C$ where $k_B$ is the Boltzmann constant, $T$ the temperature and $\eta$ the solvent viscosity.[22,23]

*Zeta potential:* Laser Doppler velocimetry was used to carry out the electrokinetic measurements of electrophoretic mobility and zeta potential with the Zetasizer Nano ZS equipment (Malvern Instruments, Worcestershore, United Kingdom). Light scattering and zeta potential measurements were performed in triplicate at 25 °C after an equilibration time of 120 s.[24]

*Transmission electron microscopy coupled with X-ray dispersive energy:* The BWP nano-sized fraction was characterized by Transmission Electron Microscopy (TEM) using a JEOL 2010 microscope (JEOL Ltd, Tokyo, Japan) operating at 200 kV with a LaB6 filament and equipped with an Orius CCD camera (Gatan inc., Warrendale, PA, USA). Energy dispersive X-ray spectroscopy (EDX, PGT) investigations were performed to determine the elements from a single nanoparticle. To avoid parasitic contribution from copper on the EDX spectra, a molybdenum TEM grid and a beryllium sample holder were used. TEM image analysis was performed with ImageJ software, and the size distributions were obtained from $n$ = 200 particles. The distributions were adjusted using a log-normal function, leading median diameter and dispersity.[25]

**Toxicological studies**

*Cell culture:* The human lung epithelial cell line Calu-3 has been shown to be relevant targets for the study of airborne particles on the respiratory tract.[26] Calu-3 cells (ATCC, HTB-55$^{TM}$, Manassas, VA, USA) were grown at 37°C under 5% $CO_2$ atmosphere in EMEM medium (Sigma Aldrich, Saint-Quentin Fallavier, France) supplemented with 10% foetal bovine serum (FBS, Eurobio, Les Ulis, France), 1% non-essential amino acids (Sigma Aldrich). Cells were trypsinized once a week (Trypsin-EDTA, Sigma Aldrich). Passages 25 to 29 were used.

*WST-1 assays:* Calu-3 cells were seeded at 40000 cells per well in 96-well plates. On the second day, cells were exposed for 24 h to BWPs and to its nano-sized fraction at increasing concentrations between 1 to 100 µg $cm^{-2}$, corresponding to weight concentrations 3 to 300 µg $mL^{-1}$. The cell viability was measured using the WST-1 assay performed using Cell Proliferation Reagent WST-1 (Roche, Basel, Switzerland) according to the manufacturer protocol. The experiments were performed in duplicates following the protocols developed by Vietti and coworkers.[27] After treatment removal, cells were rinsed twice with HBSS (Hank's Balanced Salt Solution, Invitrogen, Waltham, USA). A first plate was incubated with WST-1 dilution (1:10) in cell culture medium for 30 min, whereas the second one was treated with Triton X100 (0.1 wt. %) for 15 min before WST-1 addition. This second plate was used to evaluate potential interferences of adsorbed/internalized particles with WST-1. The absorbance signal reflecting the cell viability (mitochondrial activity) was measured at 450 and 630 nm with an ELx808 microplate reader (Biotek Instruments, Winooski, United States). The optic density at 630 nm was subtracted from the value at 450 nm to obtain the final signal.

*Reactive Oxygen Species (ROS):* Calu-3 cells were seeded at 40000 cells per well in 96-well plate. Cells were first rinsed with HBSS then loaded with 25 µM solution of CM-$DCFH_2DA$ (Molecular Probes, Eugene, United States) in HBSS for 45 min at 37°C. Following a washing step with HBSS, cells were incubated with treatment during 0.5, 1, 2 and 4 h. Fluorescence was measured with FlexStation® 3 multi-mode microplate reader (Molecular Devices, Sunnyvale, United States) at 485/535nm. Fluorescence background at 535 nm was subtracted to the fluorescence signal at 485 nm. Positive control consisting in tert-butyl hydroquinone (TBHQ) and $H_2O_2$ dilutions at 50 µm in cell culture medium was used according to manufacturer protocol.

*Cytokine release:* Calu-3 cells were seeded and treated in the same conditions as above. Supernatant were recovered and centrifuged (3000g, 4 °C) after treatment for quantification of IL-6, IL-8 and TNFα release in cell culture medium by ELISA (Enzyme-Linked Immunosorbent Assay). Absorbance signal reflecting cytokine release was measured at 450 and 630 nm with a ELx808 microplate reader (Biotek Instruments, Winooski, United States). Optic background density (OD) at 630 nm was subtracted to OD at 450 nm to obtain the final signal.

# Results and Discussion

**Brake wear particle characterization**

The BWP samples collected from passenger vehicles and from an automotive brake dynamometer have the appearance of a fine black powder (Fig. 1a). The BWPs were characterized using different techniques including thermogravimetry, X-ray fluorescence, optical and scanning electron microscopy (Fig. 1). Fig. 1b shows thermogravimetry data obtained between room temperature and 700 °C. The mass variation reveals an initial 6% decrease that is attributed to the loss of bound water around 100 °C and to the degradation of the carbon-based binder, resin and additives. This amount is lower than expected, as binders in brake pads are known to account for 20 – 40 % of the lining material. This result could be explained by the heat generated during braking that leads to high temperature and to the thermal degradation of the carbonaceous materials.[9,13,14]

At the micrometer length scale, the particle size distribution was studied using phase-contrast optical microscopy. Fig. 1c highlights the presence of contrasted particles in the range 200 nm – 10 µm, with a distribution peaked at 1.47 µm (median value) and a dispersity of 0.47 (Supplementary Information S1). Particles larger than 10 µm were also found but in a smaller proportion. The insets in Fig. 1c display close-up view of particles or aggregates of increasing size and contrast. X-ray fluorescence spectrometry was used to determine BWP elemental composition (Fig. 1d). The technique reveals 18 different elements at a level above 0.1%, including for the most prominent ones iron (Fe, 29%), copper (Cu, 17%), silicon (Si, 13%), aluminum (Al, 11%), zinc (Zn, 8%) and sulfur (S, 7%). The proportions and compositions obtained are consistent with those known for low and semi-metallic brake pads and drums typical for the European market.[7-9,11,19] They are also in line with the data received for the dynamometer test bench BWPs, indicating the weak impact of environmental driving conditions on the studied specimens (Supporting Information S2). The outcomes confirm the presence of metals such as iron, copper and zinc known to induce toxicity to living cells. We also note that antimony was not found from the elemental results and that barite was not used as filler in the brake pad formulations studied. Scanning electron microscopy (Fig. 1e) reveals that at the micro- and nanoscale the BWPs display a wide range of sizes and morphologies.[19,20,28-30] The left-hand side images show fine particles with sharp edges and rough surfaces, as well as numerous pittings arising from material removal (arrows). Micron size objects appear as shavings and fragments generated by metallic surface friction.[7] At a sub-micron scale (right hand-side panels), the surfaces have either platelet packed solid structures or show a large number of smaller particles associated with the background debris. From these data, it is not clear whether the sub-micron particles adsorb during the braking and emission events or later during sample collection and treatment. For the separation and toxicity assays presented in the next sections, emphasis will be put on the BWPs from passenger vehicles. Data from the dynamometer test bench can be found in the Supplementary Information S2 and S3.

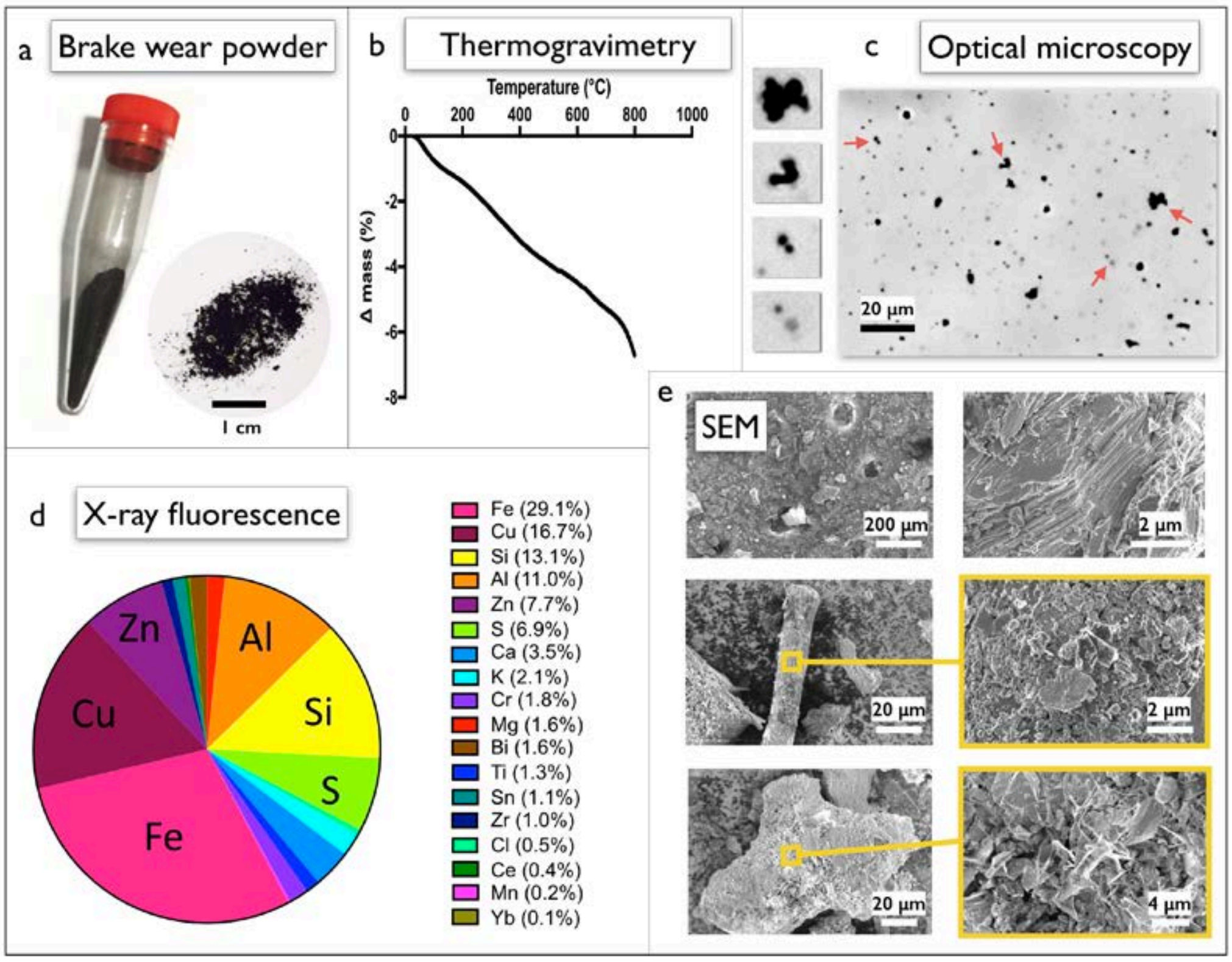

**Figure 1: Brake wear particle characterization**

*a) BWP sample obtained from a motor vehicle testing center operating on commercial transport vehicles. BWP composition was characterized using thermogravimetry analysis measured in helium atmosphere (b) and X-ray fluorescence for elemental analysis (d). To evaluate the size and morphology of the wear particles optical (c) and scanning electron (e) microscopy were performed. The insets in c) are 7×8 µm$^2$ fields.*

## Sonication and separation of brake wear particles in water

SEM has shown the existence of sub-100 nm particles, and an important question pertaining to the toxicity assessment is that of the nano-sized fraction.[3] Sonication is a well-known technique used to break particle aggregates as well as to improve particle dispersability. BWP were dispersed in DI-water at 1 mg mL$^{-1}$, vortexed and sonicated as a function of the delivered sonication energy $DSE$[31,32] under conditions described in the Materials and Methods section. The suspensions were then filtered through a 0.45 µm membrane and characterized by static and dynamic light scattering. Fig. 2a show the scattered intensity $I_S$ (upper panel) and the hydrodynamic diameter $D_H$ (lower panel) against the $DSE$ for filtered and non-filtered BWP dispersions.[22] Prior to sonication, the dispersion exhibits a strong scattering intensity and diameters larger than 1 µm, in agreement with the optical and electron microscopy results. The filtered solution in contrast has a much weaker scattering, about 400 times smaller, indicating that as anticipated filtration did remove most suspended particles. With increasing $DSE$, the scattering intensity increases and displays a saturation plateau. In parallel the diameter decreases and levels off at 165 nm. The corresponding autocorrelation functions $g^{(2)}(t)$ obtained from dynamic light scattering exhibit a quasi-exponential decay associated with a unique relaxation mode[23] and a distribution that refines with increasing $DSE$ (Fig. 2b). Note that at the highest input energy, the intensity for filtered and unfiltered dispersions are close to each other, suggesting that sonication was able to redisperse most particles. With a zeta potential of -26 mV, electrokinetic measurements confirmed that brake wear nanoparticles are negatively charged and that their stabilization is ensured by electrostatics (Fig. 2c). We also found that the surface charge does not depend on the sonication energy. The previous experiments were repeated in presence of bovine serum albumin (BSA) proteins at 1 g $L^{-1}$ to evaluate whether adsorbed macromolecules could improve the dispersion stability. The figures in Supplementary Information S4 show that BSA has no significant effect on the particle dispersability, even at long term (> 1 month). To determine the proportion of nanoparticles in the BWP samples, the dispersions were evaporated in an oven and the residual dry matter weighed, revealing that 26 ± 1% of the initial BWP is indeed made up of nanoparticles. This percentage is consistent with the static light scattering results of Fig. 2a.

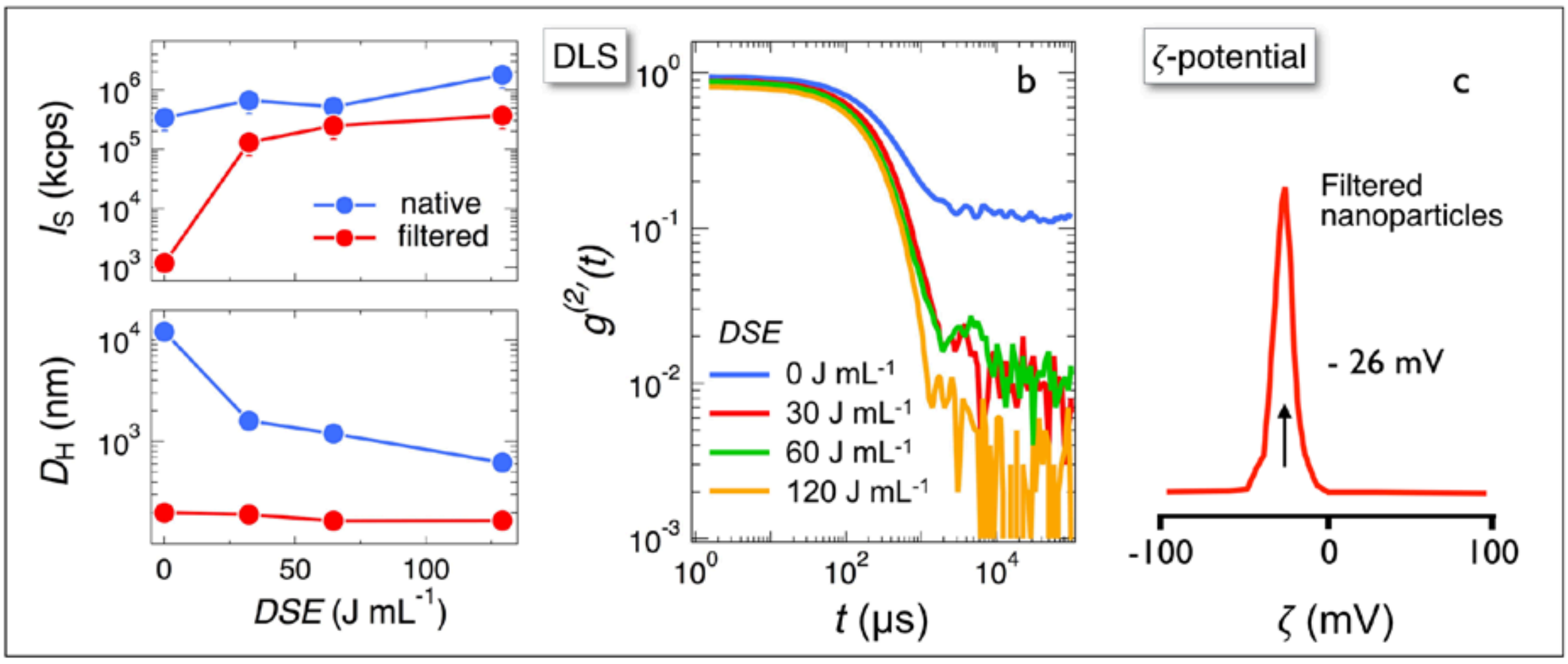

**Figure 2: Characterization of brake wear particles before and after sonication and filtration**

*a) Scattering intensity (upper panel) and hydrodynamic diameter (lower panel) measured by light scattering on 1 mg $mL^{-1}$ sonicated BWP dispersions as a function of the delivered sonication energy DSE.[31,32] In the configuration used, a one-minute sonication translates into a total input energy of 30 J $mL^{-1}$. b) Second-order autocorrelation functions of the scattered intensity for the filtered solutions in a). With increasing DSE, the dispersity indexes (pdi) were 0.28, 0.29, 0.20 and 0.16, respectively. c) Intensity distribution as a function of the zeta potential ζ for a sonicated and filtered dispersion showing that BWP nanoparticles are negatively charged.*

## Brake wear nano-sized fraction: determination of size, shape and composition

The BWP nano-sized fraction was further characterized by transmission electron microscopy coupled with X-ray dispersive energy spectroscopy. As shown in Figs. 3a and 3b, nanoparticles have sizes comprised between 50 and 150 nm and a broad range of morphologies. The most abundant elements are copper and iron, in good agreement with the X-ray fluorescence data. Interestingly, EDX spectra from different brake wear nanoparticles show the presence of one dominant of metal per particle, with traces of others, indicating that the particles were generated from different parts of the pads and drums. Examples are provided in Supporting Information S5 for iron, silicon, zirconium, tin, bismuth and titanium. Fig. 3c displays a size distribution peaked at 66 nm (median value) and characterized by a dispersity of 0.50. Studied at different times after sonication, the BWP nanoparticles were found to be stable over time, with hydrodynamic diameters remaining constant over a month (Fig. 3d). In cell culture medium it was found that the nano-sized fraction slowly destabilizes as a function of time, giving rise to the formation of aggregates.[33] Aggregation is here primarily due to the adsorption of proteins and/or biological molecules at the particle surfaces that modify the interaction between particles. From repulsive in DI-water, the interaction becomes attractive in presence of salt or proteins, leading to the particle destabilization.[34-36] Fig. S6 displays the autocorrelation function and the size distribution after a 24 h incubation at 100 µg $mL^{-1}$. The data in Supporting Information exhibit three relaxation modes, one associated with the initial particle size (160 nm, 12%), the second and third modes being associated with medium (320 nm, 78%) and large (1200 nm, 10%) aggregates. The percentages in parenthesis indicate here the frequency of each contribution. These results will be exploited later in the manuscript to determine the effective dose administered to the cells using the *in vitro* sedimentation, diffusion and dosimetry model recently developed.[31,32] To conclude, we combined sonication and filtration to separate the BWP nano-sized fraction from the original dust. This protocol will be now applied for *in vitro* toxicity assessment studies.

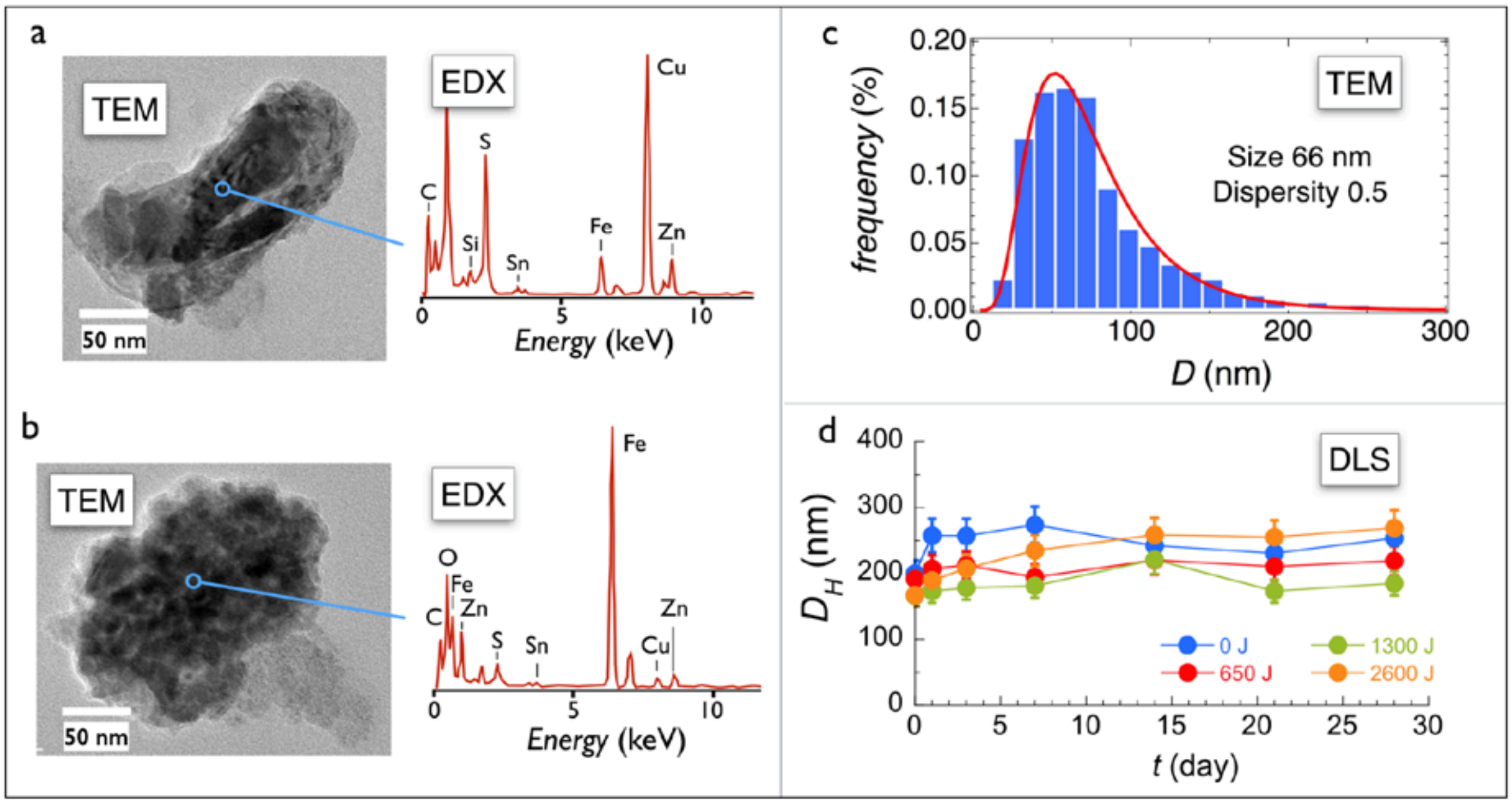

**Figure 3: Composition, size and stability of the brake wear nano-sized fraction**
*a,b) left: TEM images of brake wear nanoparticles; right: EDX spectra of the particle region shown in the blue circles. c) Size distribution determined from TEM on the filtered particles. The median value is 66 nm and the dispersity 0.50. d) Hydrodynamic diameter as a function of the time for brake wear nanoparticle dispersions at increasing sonication energies. In the four instances, the particles are stable as the diameter remains unchanged for the period.*

**Biological effects of brake wear particles and its nano-sized fraction on Calu-3 cells**

The toxicity of BWP and its nano-sized fraction was investigated on human bronchial epithelial cells Calu-3.[26] The cytotoxicity was assessed using the WST-1 mitochondrial activity assay[37] for concentrations 1 to 100 µg $cm^{-2}$. After a 24 h exposure, the cell viability was found to decrease with the applied dose for both samples (Fig. 4a). However, the BWP exhibited a stronger effect, the viability being reduced by 81% at 100 µg $cm^{-2}$, this value becoming 35% for the nanoparticle dispersion. Zhao *et al.* observed that some brake wear samples reduce the viability of A549 alveolar cells, and that the decrease was comparable to ours.[19] Exposing the same cell line to an aerosol of freshly produced BWP however Gasser *et al.* did not notice any significant effect. This apparent discrepancy could be due to the different plating techniques used, cells being grown on inserts providing a denser epithelium compared to the conditions in plates, leading to lower particle dose per cell.[18]

BWP being mainly composed of metal compounds (as shown in Fig. 1), this loss of viability may be related to the oxidative stress arising from the imbalance between intracellular oxidants and anti-oxidant molecules.[38] Metallic ions can generate ROS through the Fenton reaction for transition metals and by depleting thiol anti-oxidant molecules such as glutathione.[2] The intracellular redox state was assessed using the DCFH-DA probe that exhibits green fluorescence upon oxidation. A time-course study over 4 hours revealed a time dependent increase of the DCF fluorescence in cells treated with either BWP or with its nano-sized fraction. The fluorescence signal remains however lower than that of positive $H_2O_2$ and T-BHQ controls (Supporting Information S7). Fig. 4b displays the ROS production as a function of the dose following a 4 h incubation together with the controls. Both BWP and nanoparticles exhibit a similar behavior, with a nearly two-fold increase at the highest dose. These results are in fair agreement with those of the literature for A549 cells treated in the same conditions.[18]

The ability of BWP to induce a pro-inflammatory response was characterized by evaluating the release of the IL-6, IL-8 and TNF$\alpha$ cytokines after a 24 h exposure (Fig. 4c). While for the BWP IL-8 and TNFα releases were not significantly modified (Fig. S8), the IL-6 release was found to be dose dependent, with a deviation seen already at 10 μg cm$^{-2}$, *i.e.* at the concentration where cell viability is reduced by 50%. At the highest dose, the IL-6 release shows a 5-fold increase compared to the control. The situation is different for the nano-sized fraction for which IL-6 and IL-8 signals remain dose independent and similar to control. The toxicity, ROS production and pro-inflammatory response results in Fig. 4 reveal an overall lower toxicity for the brake wear nanoparticles compared to that of the initial fraction. We have checked whether this extra toxicity could be due to the soluble fraction that is present in the BWP dispersions (and not in the nanoparticle dispersions) and may contain harmful metallic ions. The cytotoxicity of the soluble fraction was evaluated by WST-1 assay and revealed a weak decrease of viable cells by 10% at the highest concentration, suggesting that the solubilized ions do not contribute significantly to the observed phenomenon (Fig. S9).

From the above results, the nano-sized fraction appears hence to be less toxic than BWP. However the comparison was made on the basis of the nominal concentrations and not on the effective delivered concentrations. To examine this issue, we have applied the *In vitro* Sedimentation, Diffusion and Dosimetry (ISDD) model recently developed by DeLoid and coworkers to improve dose metrics methodologies.[31,32] According to the ISDD model, the doses administered to the cells depend primarily on the particle sizes and on their densities. For brake wear nanoparticles we based our calculations on the light scattering data of Fig. S5, as suggested by the model. For the BWP, we used the size determined from optical microscopy (Fig. S1), the light scattering data being not reliable for micron size particles. The results of the ISDD simulations shown in Supporting Information S10 include the time dependencies of the deposited fraction and bottom concentration, and the concentration gradients in the well. It was found that the micron size BWP sediment rapidly and that the delivered dose reaches 79% after 24 h of exposure. For the nanoparticle in contrast, the dose delivered to cells was estimated to reach 7 % of the nominal dose in the same conditions. Considering the effective doses, the mitochondrial activity results of Fig. 4 can be re-interpreted. At 50 μg cm$^{-2}$ for instance, the effective dose for nanoparticles at 24 h is around 4 μg cm$^{-2}$ and the cell viability 70%. For the BWP, an effective dose of 4 μg cm$^{-2}$ corresponds to a nominal dose of 5 µg cm$^{-2}$, for which the cell viability is 60%, so close to that of the previous value. In conclusion, the differences seen in the cell viability for regular BWP and for its nano-sized fraction are mainly due to dosimetric effects, demonstrating that BWPs are toxic, but the nano-sized fraction does not display extra toxicity. Similar findings have been identified for the pro-inflammatory response, whereas for the ROS production nanoparticles exhibit a more pronounced effect (Supplementary Information S11). Altogether these data suggest that under the conditions of the study, the brake wear macro- and nanoparticles exhibit similar cytotoxicity. We can speculate that the underlying mechanisms leading to cell death could be different, as indicated by the difference in ROS production.

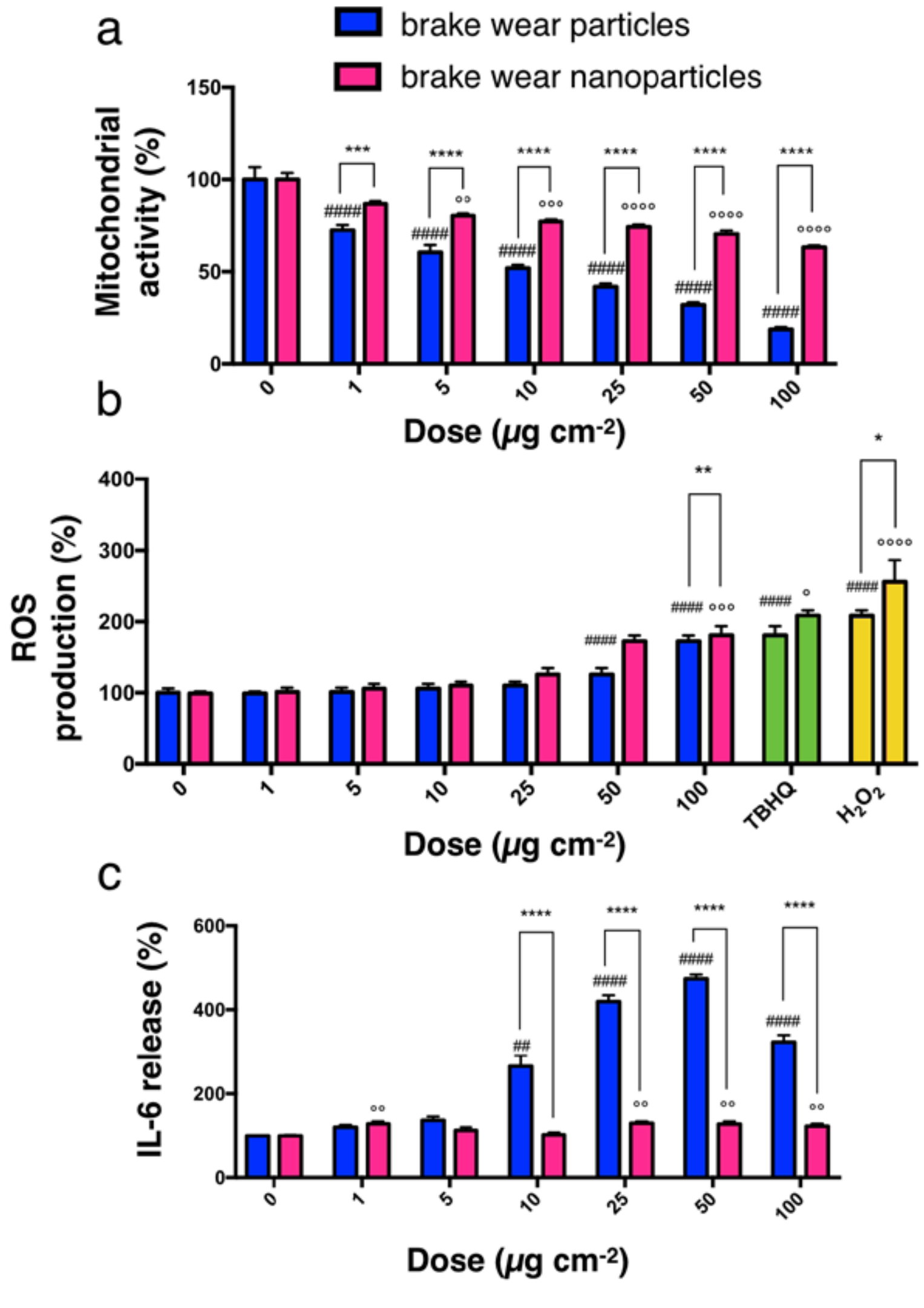

**Figure 4: Biological effects of brake wear particles and its nano-sized fraction on Calu-3 cells**

*Calu-3 cells were exposed to BWPs and to its nano-sized fraction at concentrations ranging from 1 to 100 µg cm$^{-2}$ (corresponding to effective weight concentrations in the supernatant of 3 to 300 µg mL$^{-1}$). a) Mitochondrial activity reflecting cell viability were assessed using WST-1 assay after a 24 h of exposure. b) Reactive oxygen species production was detected by DCFH probe oxidation after 4 h exposure. c) IL-6 cytokine release measured by ELISA was used to characterize the pro-inflammatory response after a 24 h exposure. Statistical analysis was performed using One-way ANOVA and Dunett's post-test to compare particles effects with no exposure. The symbols (#) and (°) refer to the*

*BWP and to its nano-sized fraction respectively. Two-way ANOVA complemented by Holm-Sidak post test was performed to compare the BWP size effects (*).*

# Conclusion

We have shown that BWPs emitted from low-metallic brake pads used in passenger cars and coming from two different sources have typical sizes in the micron range and contain metallic compounds, such as iron, copper, silicon, aluminum and zinc under crystalline and amorphous forms. Scanning electron microscopy indicates that the initial BWPs appear as fragments or debris made from a large number of sub-micron particles. To isolate these sub-micron particles we set up a combination of sonication, centrifugation and filtration techniques. The first significant conclusion that emerges from this work is that the nano-sized fraction represents around 26% by mass of the initial brake wear dust. The nanoparticles are also broadly distributed, with size comprised between 50 and 400 nm. Their toxicity was assessed on human bronchial epithelial cells. The second important result is that BWPs display short-term toxicity and we do not find any enhanced effect from the nano-sized fraction. These results show finally the interest of isolating this fraction of brake wear dust and characterizing its *in vitro* behavior, in particular in the cell culture medium for accurate dosimetry and toxicity assessment.

# Acknowledgments

The authors thank DIM (Domaine d'Intérêt Majeur) Nano-K of Region Île-de-France for funding C. Puisney Ph. D. thesis at the University Paris-Diderot. ANR (Agence Nationale de la Recherche) and CGI (Commissariat à l'Investissement d'Avenir) are acknowledged for their financial support of this work through Labex SEAM (Science and Engineering for Advanced Materials and devices) ANR 11 LABX 086, ANR 11 IDEX 05 02. We acknowledge the ImagoSeine facility (Jacques Monod Institute, Paris, France), and the France BioImaging infrastructure supported by the French National Research Agency (ANR-10-INSB-04, « Investments for the future »). This research was supported in part by the Agence Nationale de la Recherche under the contract ANR-13-BS08-0015 (PANORAMA) and ANR-12-CHEX-0011 (PULMONANO).

# Conflicts of interest

There are no conflicts to declare.

# References

1. R. D. Brook, S. Rajagopalan, C. A. Pope, J. R. Brook, A. Bhatnagar, A. V. Diez-Roux, F. Holguin, Y. L. Hong, R. V. Luepker, M. A. Mittleman, A. Peters, D. Siscovick, S. C. Smith, L. Whitsel and J. D. Kaufman, *Circulation*, 2010, **121**, 2331-2378.
2. M. Valko, K. Jomova, C. J. Rhodes, K. Kuca and K. Musilek, *Arch. Toxicol.*, 2016, **90**, 1-37.
3. V. Stone, M. R. Miller, M. J. Clift, A. Elder, N. L. Mills, P. Moller, R. P. Schins, U. Vogel, W. G. Kreyling, K. A. Jensen, T. A. Kuhlbusch, P. E. Schwarze, P. Hoet,

A. Pietroiusti, A. De Vizcaya-Ruiz, A. Baeza-Squiban, C. L. Tran and F. R. Cassee, *Environ. Health Persp.*, 2016, DOI: 10.1289/EHP424.
4. F. R. Cassee, M. E. Heroux, M. E. Gerlofs-Nijland and F. J. Kelly, *Inhal. Toxicol.*, 2013, **25**, 802-812.
5. J. H. J. Hulskotte, H. A. C. D. van der Gon, A. H. Visschedijk and M. Schaap, *Water Sci. Technol.*, 2007, **56**, 223-231.
6. N. Hooftman, L. Oliveira, M. Messagie, T. Coosemans and J. Van Mierlo, *Energies*, 2016, **9**, 84.
7. T. Grigoratos and G. Martini, *Environ. Sci. Poll. Res.*, 2015, **22**, 2491-2504.
8. J. Kukutschova, P. Moravec, V. Tomasek, V. Matejka, J. Smolik, J. Schwarz, J. Seidlerova, K. Safarova and P. Filip, *Environ. Pollut.*, 2011, **159**, 998-1006.
9. J. Kukutschova, V. Roubicek, M. Maslan, D. Jancik, V. Slovak, K. Malachova, Z. Pavlickova and P. Filip, *Wear*, 2010, **268**, 86-93.
10. A. Thorpe and R. M. Harrison, *Sci. Total Environ.*, 2008, **400**, 270-282.
11. P. G. Sanders, N. Xu, T. M. Dalka and M. M. Maricq, *Environ. Sci. Technol.*, 2003, **37**, 4060-4069.
12. B. D. Garg, S. H. Cadle, P. A. Mulawa, P. J. Groblicki, C. Laroo and G. A. Parr, *Environ. Sci. Technol.*, 2000, **34**, 4463-4469.
13. A. Iijima, K. Sato, K. Yano, M. Kato, K. Kozawa and N. Furuta, *Environ. Sci. Technol.*, 2008, **42**, 2937-2942.
14. T. P. Newcomb, *Wear*, 1980, **59**, 401-407.
15. A. Nel, T. Xia, L. Madler and N. Li, *Science*, 2006, **311**, 622-627.
16. G. Oberdörster, E. Oberdörster and J. Oberdörster, *Environ. Health Persp.*, 2005, **113**, 823-839.
17. J. Wahlström, Ph. D. Thesis, University of Stockholm, 2011.
18. M. Gasser, M. Riediker, L. Mueller, A. Perrenoud, F. Blank, P. Gehr and B. Rothen-Rutishauser, *Part. Fibre Toxicol.*, 2009, **6**.
19. J. Zhao, N. Lewinski and M. Riediker, *Aerosol Sci. Tech.*, 2015, **49**, 65-74.
20. A. Kazimirova, P. Peikertova, M. Barancokova, M. Staruchova, J. Tulinska, M. Vaculik, I. Vavra, J. Kukutschova, P. Filip and M. Dusinska, *Environ. Res.*, 2016, **148**, 443-449.
21. B. Kowalczyk, I. Lagzi and B. A. Grzybowski, *Curr. Opin. Colloid Interface Sci.*, 2011, **16**, 135-148.
22. F. Mousseau, R. Le Borgne, E. Seyrek and J.-F. Berret, *Langmuir*, 2015, **31**, 7346-7354.
23. V. Torrisi, A. Graillot, L. Vitorazi, Q. Crouzet, G. Marletta, C. Loubat and J.-F. Berret, *Biomacromolecules*, 2014, **15**, 3171-3179.
24. E. K. Oikonomou, F. Mousseau, N. Christov, G. Cristobal, A. Vacher, M. Airiau, C. Bourgaux, L. Heux and J. F. Berret, *J. Phys. Chem. B*, 2017, **121**, 2299-2307.
25. L. Qi, J. P. Chapel, J. C. Castaing, J. Fresnais and J.-F. Berret, *Soft Matter*, 2008, **4**, 577-585.
26. I. George, S. Vranic, S. Boland, A. Courtois and A. Baeza-Squiban, *Toxicol. In Vitro*, 2015, **29**, 51-58.
27. G. Vietti, S. Ibouraadaten, M. Palmai-Pallag, Y. Yakoub, C. Bailly, I. Fenoglio, E. Marbaix, D. Lison and S. van den Brule, *Part. Fibre Toxicol.*, 2013, **10**.
28. H.-G. Namgung, J.-B. Kim, S.-H. Woo, S. Park, M. Kim, M.-S. Kim, G.-N. Bae, D. Park and S.-B. Kwon, *Environ. Sci. Technol.*, 2016, **50**, 3453-3461.
29. K. Adachi and Y. Tainosho, *Environ. Int.*, 2004, **30**, 1009-1017.
30. M. Mosleh, P. J. Blau and D. Dumitrescu, *Wear*, 2004, **256**, 1128-1134.

31. G. DeLoid, J. M. Cohen, T. Darrah, R. Derk, L. Rojanasakul, G. Pyrgiotakis, W. Wohlleben and P. Demokritou, *Nat. Commun.*, 2014, **5**, 3514.
32. G. M. DeLoid, J. M. Cohen, G. Pyrgiotakis and P. Demokritou, *Nat. Protoc.*, 2017, **12**, 355-371.
33. M. Safi, J. Courtois, M. Seigneuret, H. Conjeaud and J.-F. Berret, *Biomaterials*, 2011, **32**, 9353-9363.
34. G. Ramniceanu, B. T. Doan, C. Vezignol, A. Graillot, C. Loubat, N. Mignet and J. F. Berret, *RSC Advances*, 2016, **6**, 63788-63800.
35. A. Galimard, M. Safi, N. Ould-Moussa, D. Montero, H. Conjeaud and J. F. Berret, *Small*, 2012, **8**, 2036-2044.
36. B. Chanteau, J. Fresnais and J.-F. Berret, *Langmuir*, 2009, **25**, 9064-9070.
37. M. Delaval, W. Wohlleben, R. Landsiedel, A. Baeza-Squiban and S. Boland, *Arch. Toxicol.*, 2017, **91**, 163-177.
38. B. Crobeddu, L. Aragao-Santiago, L. C. Bui, S. Boland and A. Baeza Squiban, *Environ. Pollut.*, 2017, **230**, 125-133.

# TOC image

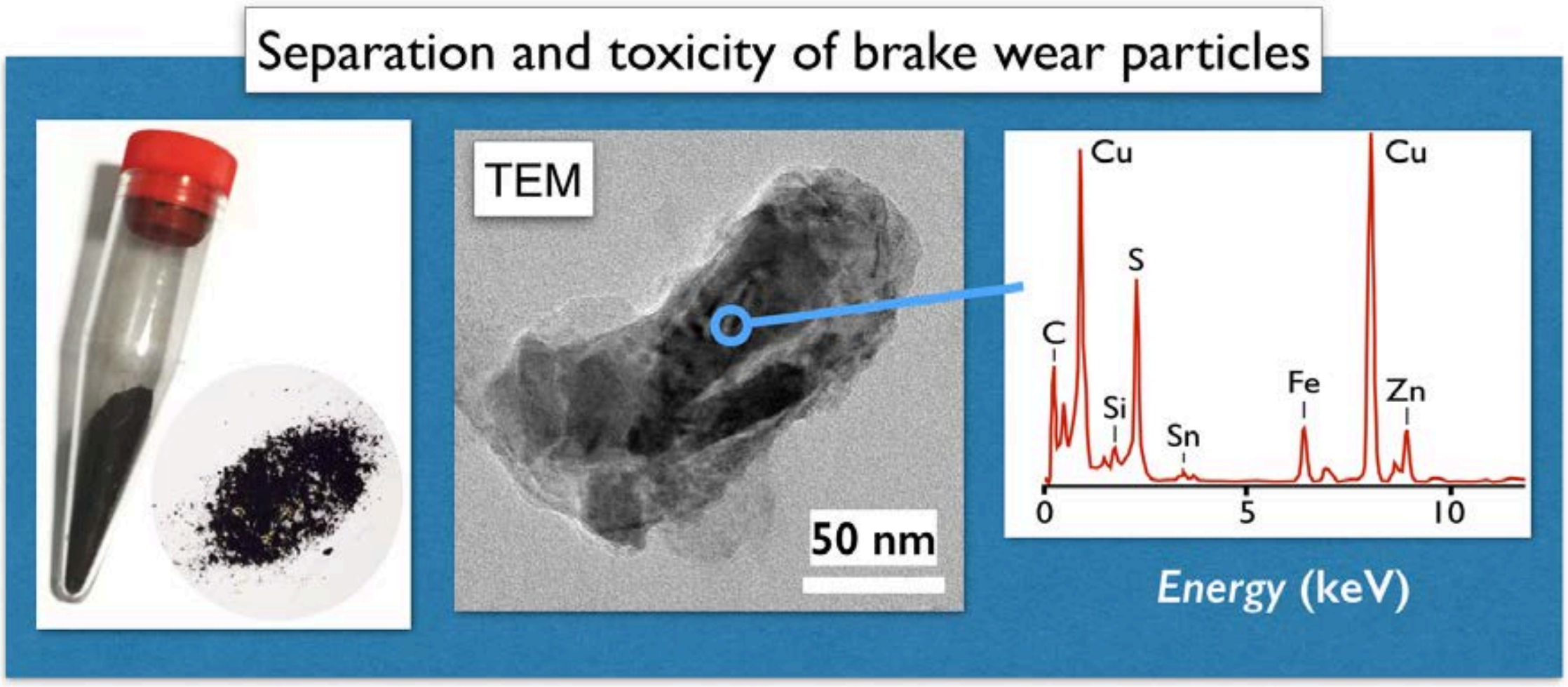

# Supporting Information

## Brake wear (nano)particle characterization and toxicity on airway epithelial cells *in vitro*

C. Puisney[1,2*], E.K. Oikonomou[2], S. Nowak[3], A. Chevillot[3], S. Casale[4], A. Baeza-Squiban[1*] and J.-F. Berret[2*]

### Outline

Corresponding authors: chloe.puisney@univ-paris-diderot.fr, baeza@univ-paris-diderot.fr, jean-francois.berret@univ-paris-diderot.fr

## Supporting Information Figure S1 – Size distribution from BWP found on commercial transport vehicles

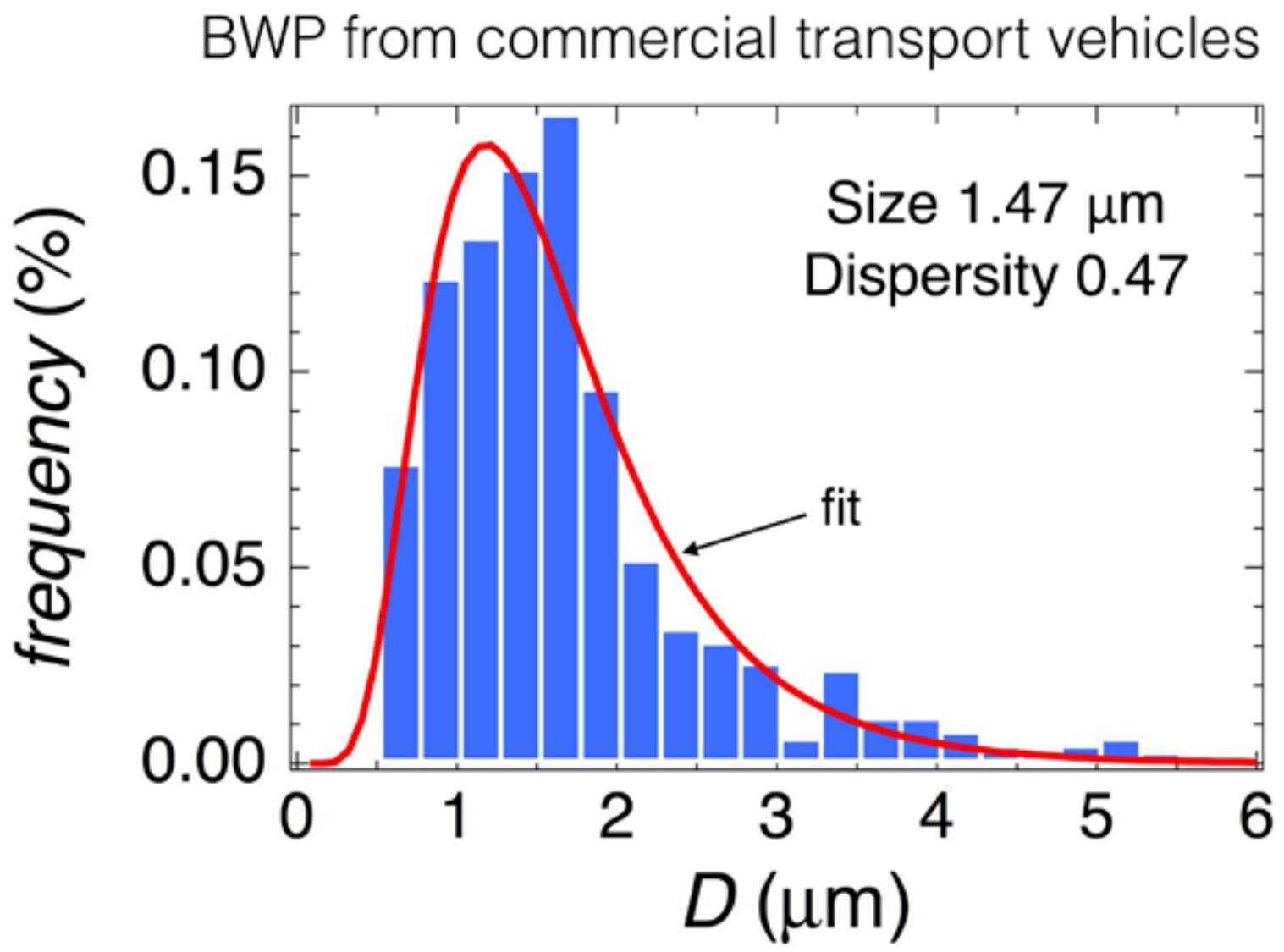

***Figure S1:*** *Distribution of particle detected by optical microscopy obtained from a motor vehicle testing center operating on light duty vehicles. The fitting curve is given by a log-normal function.*

## Supporting Information Figure S2 – Characterization of brake wear powder

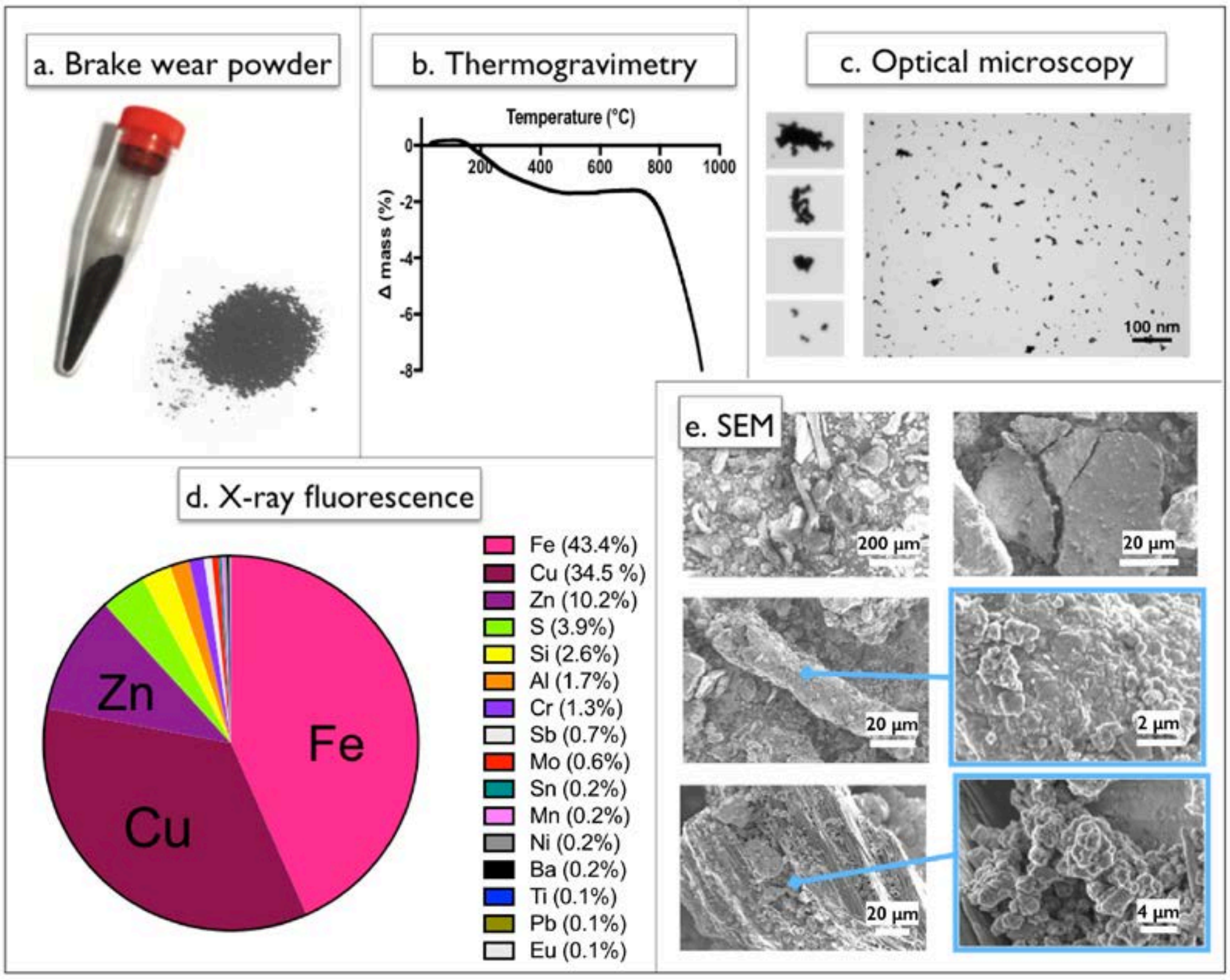

***Figure S2: Characterization of brake wear powder originating from an automotive brake dynamometer***

*a) Brake wear powder sample obtained from an automotive brake dynamometer for pad and drum testing. BWP composition was characterized using thermogravimetry analysis (b) and X-ray fluorescence for elemental analysis (d). Note that for the sample coming from the motor vehicle testing center, the iron and copper proportions were* 29% and 17% respectively, whereas the values here are 43 and 34 %. *Moreover, this brake pad formulation contained Sb and Mo, most probably in form of sulfides (since content of S is 3.9%). Content of Ba (0.2%) suggests that Ba was not used as a filler. To evaluate the size and morphology of the wear particles optical (c) and scanning electron (e) microscopy were performed. The insets in c) are 24×40 µm$^2$ fields.*

## Supporting Information Figure S3 – Size distribution from BWP collected from an automotive brake dynamometer

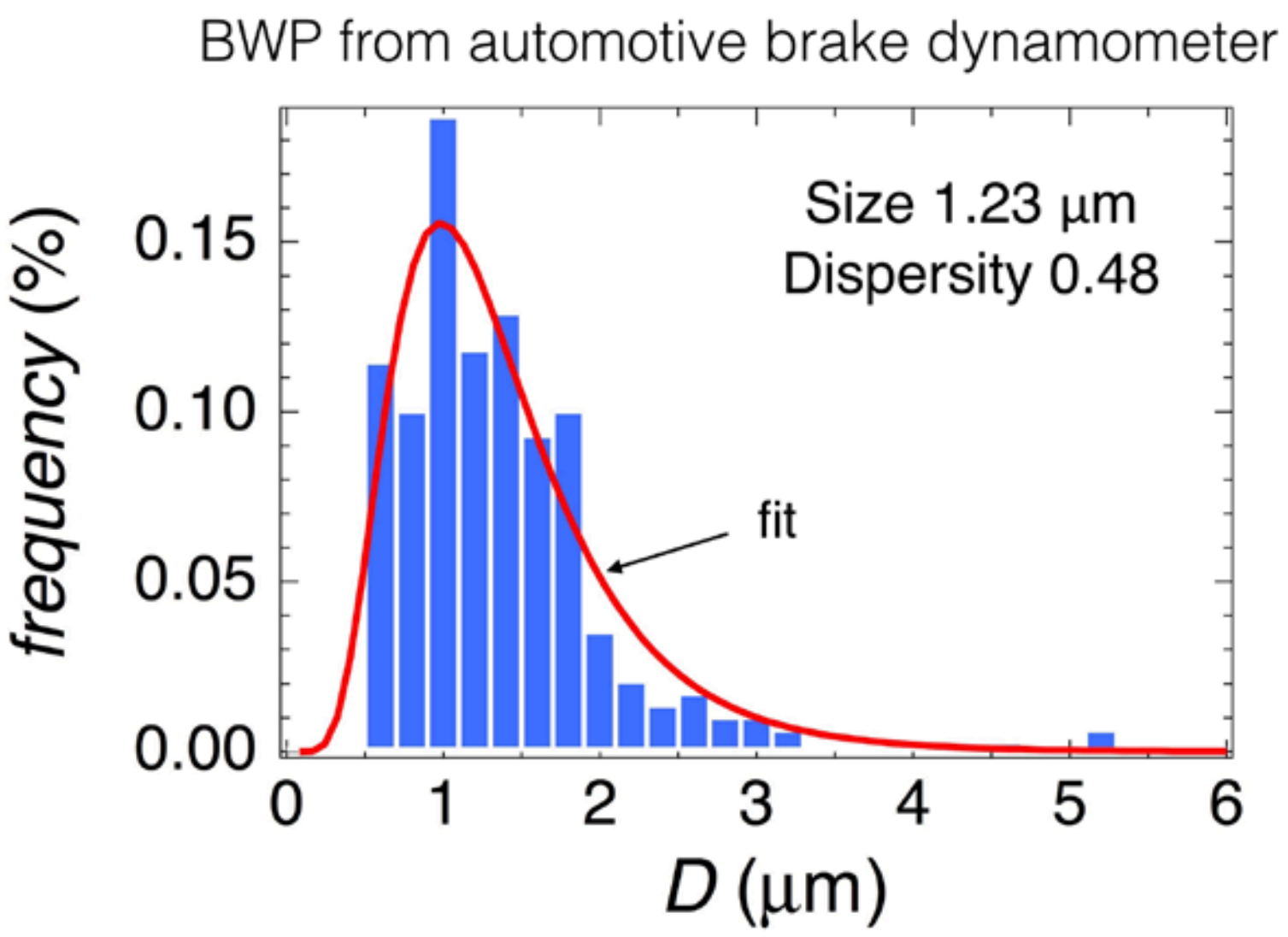

***Figure S3:*** *Distribution of particle detected by optical microscopy obtained from an automotive brake dynamometer for pad and drum testing. The fitting curve is given by a log-normal function.*

## Supporting Information Figure S4 – Light scattering data for BWP dispersed and sonicated with proteins

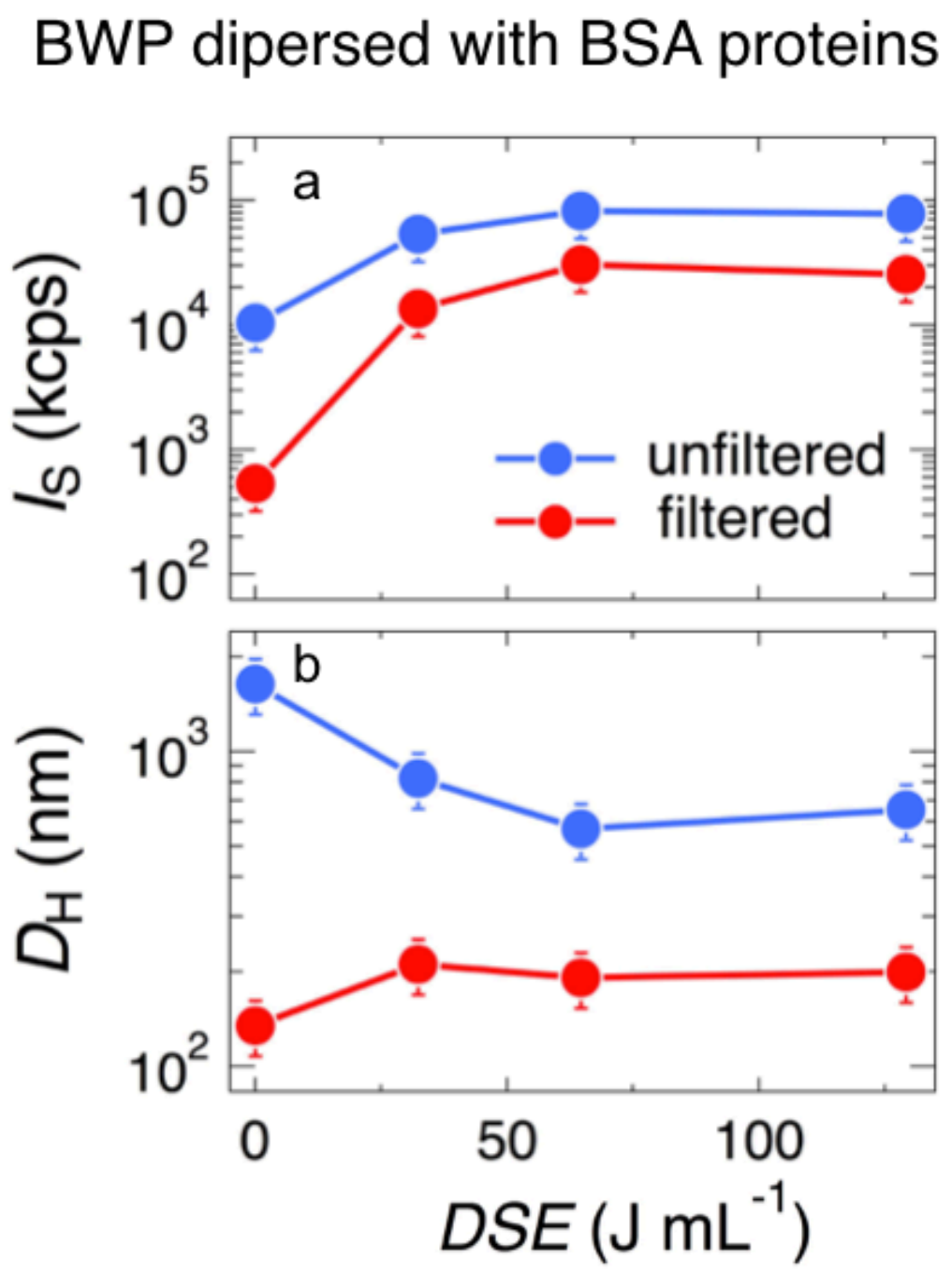

***Figure S4:*** a) Scattering intensity and b) hydrodynamic diameter measured by light scattering of BWP particles dispersed in a 1 g L$^{-1}$ BSA solution as a function of the sonication delivered sonication energy $DSE$. In the configuration used, a one-minute sonication translates into a total input energy of 30 J mL$^{-1}$.

## Supporting Information Figure S5 – Dynamic light scattering in cell culture medium (EMEM)

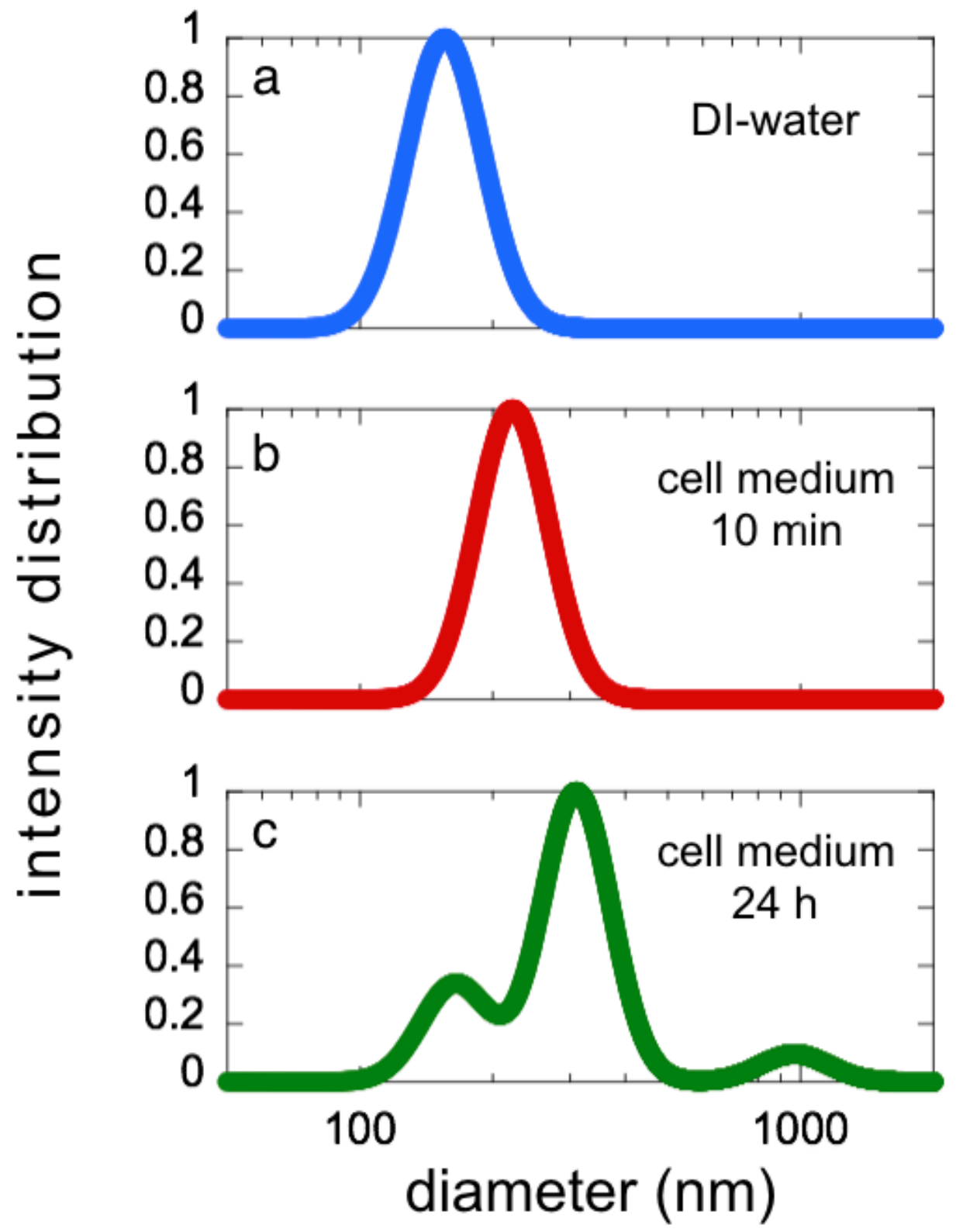

***Figure S5:*** *Intensity distributions for the BWP nano-sized fraction in DI-water (a) and in cell culture medium (EMEM) at 10 min (b) and 24 h (c). The maxima at 24 h are at 170 nm, 320 nm and 1000 nm.*

## Supporting Information Figure S6 – Complementary TEM images and EDX spectra from brake wear nanoparticles

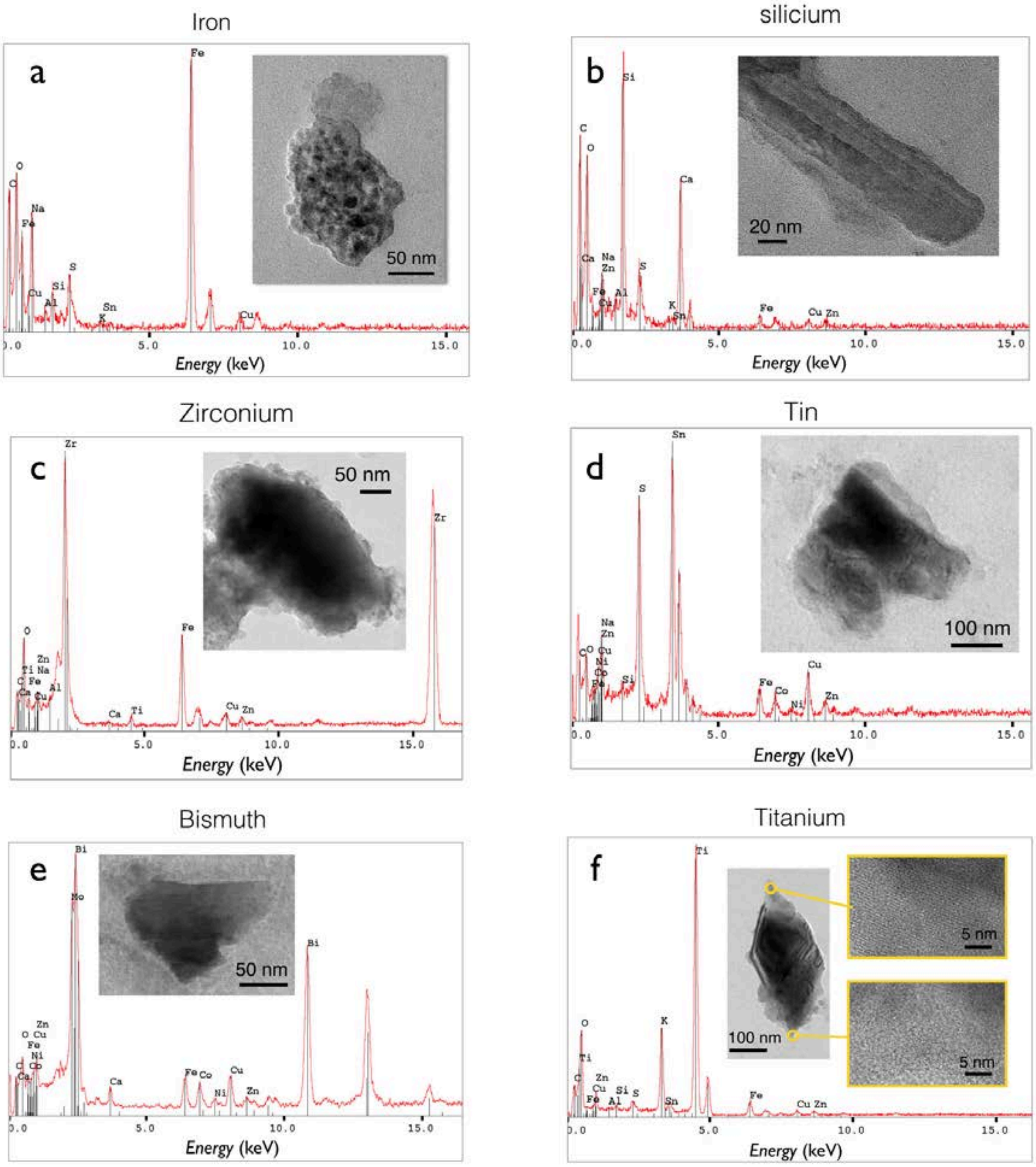

***Figure S6:*** a-f) EDX spectra from different brake wear nanoparticles showing that elemental composition underlines the presence of only one type of atoms per particle, indicating that the particles are generated from different parts of the braking component. Examples of particles containing iron (a), silicon (b), zirconium (c), tin (d), bismuth (e) and titanium (f) are provided.

## Supporting Information Figure S7 - Reactive oxygen species production in Calu-3 cells

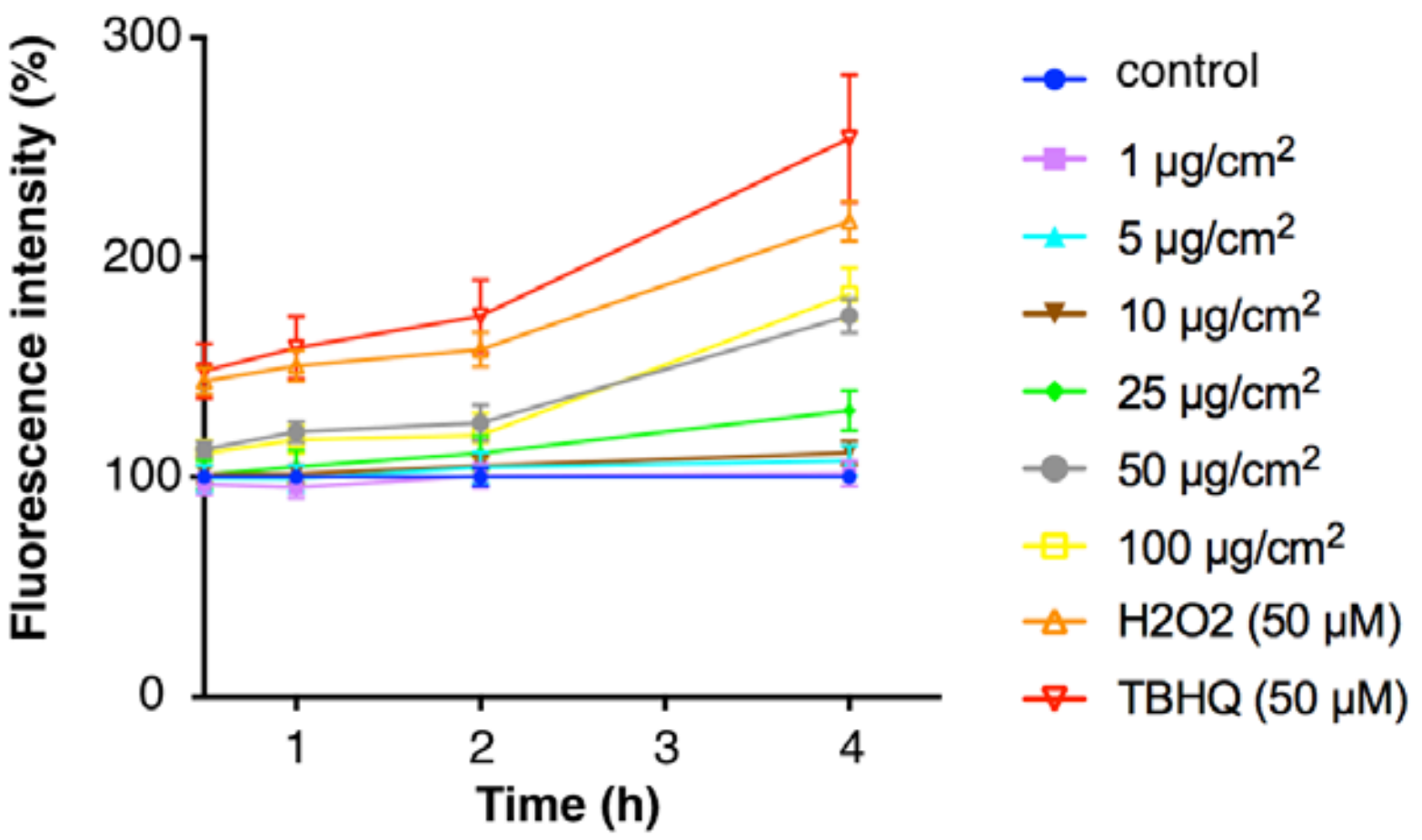

***Figure S7:*** *Reactive oxygen species production was detected by DCFH probe fluorescence in Calu-3 cells exposed for 4 h to BWP from 1 to 100 µg cm*$^{-2}$*.* $H_2O_2$ *and TBHQ were used as positive controls*

## Supporting Information Figure S8 – IL-6 and TNF$\alpha$ release by Calu-3 cells exposed to BWP or its nanosize fraction

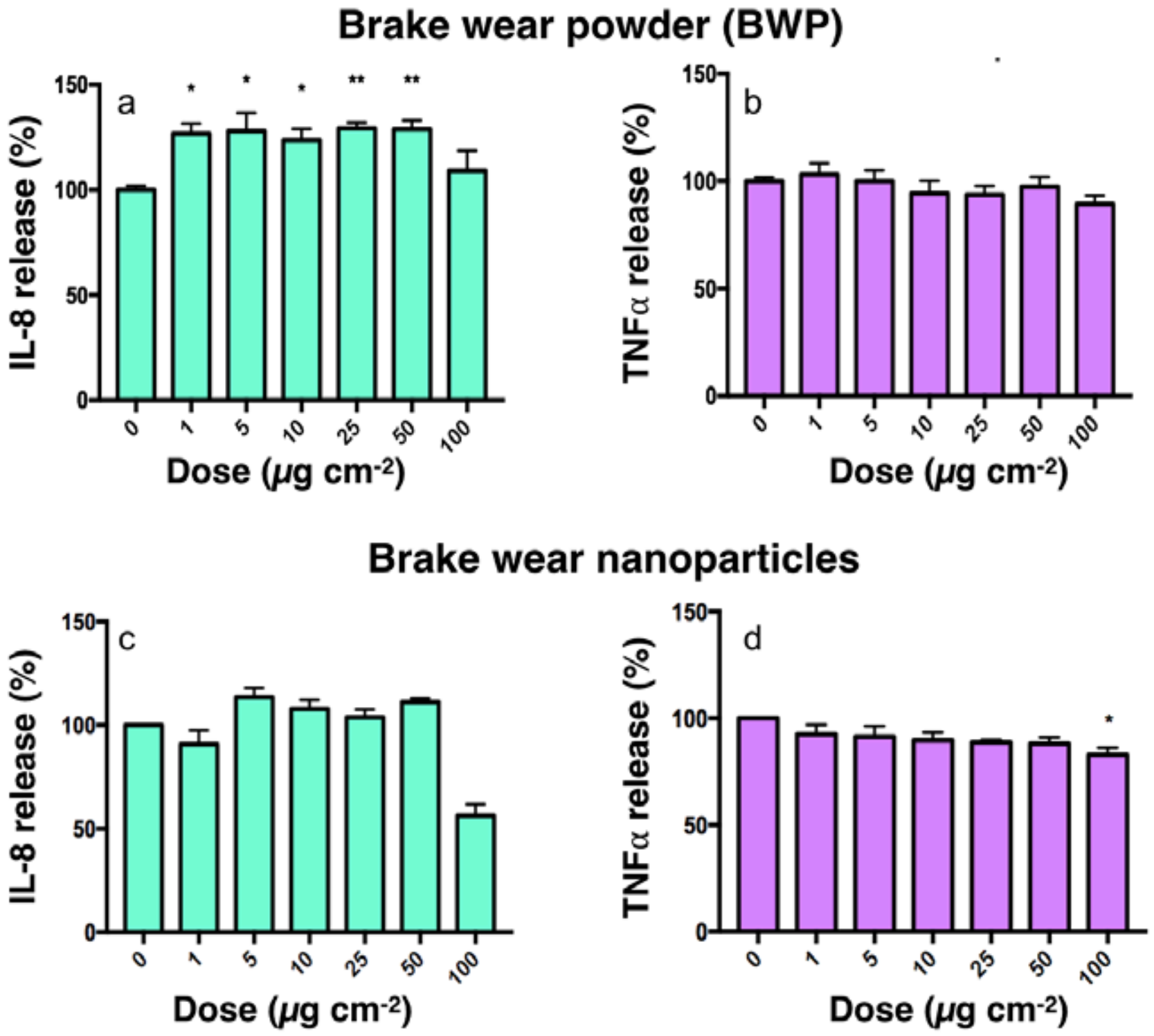

***Figure S8:*** *IL-8 and TNF$\alpha$ cytokine release by Calu-3 cells exposed for 24 h to BWP and its nanosize fraction at different concentrations. Cytokine measured by ELISA was used to characterize the pro-inflammatory response after a 24 h exposure. Statistical analysis was performed using One-way ANOVA and Dunett's post-test to compare particles effects with no exposure. The symbols (#) and (°) refer to the BWP and to its nanosize fraction respectively. Two-way ANOVA complemented by Holm-Sidak post-test was performed to compare the BWP size effects (*).*

## Supporting Information Figure S9 - Cell viability of Calu-3 cells exposed either to particles or the soluble fraction

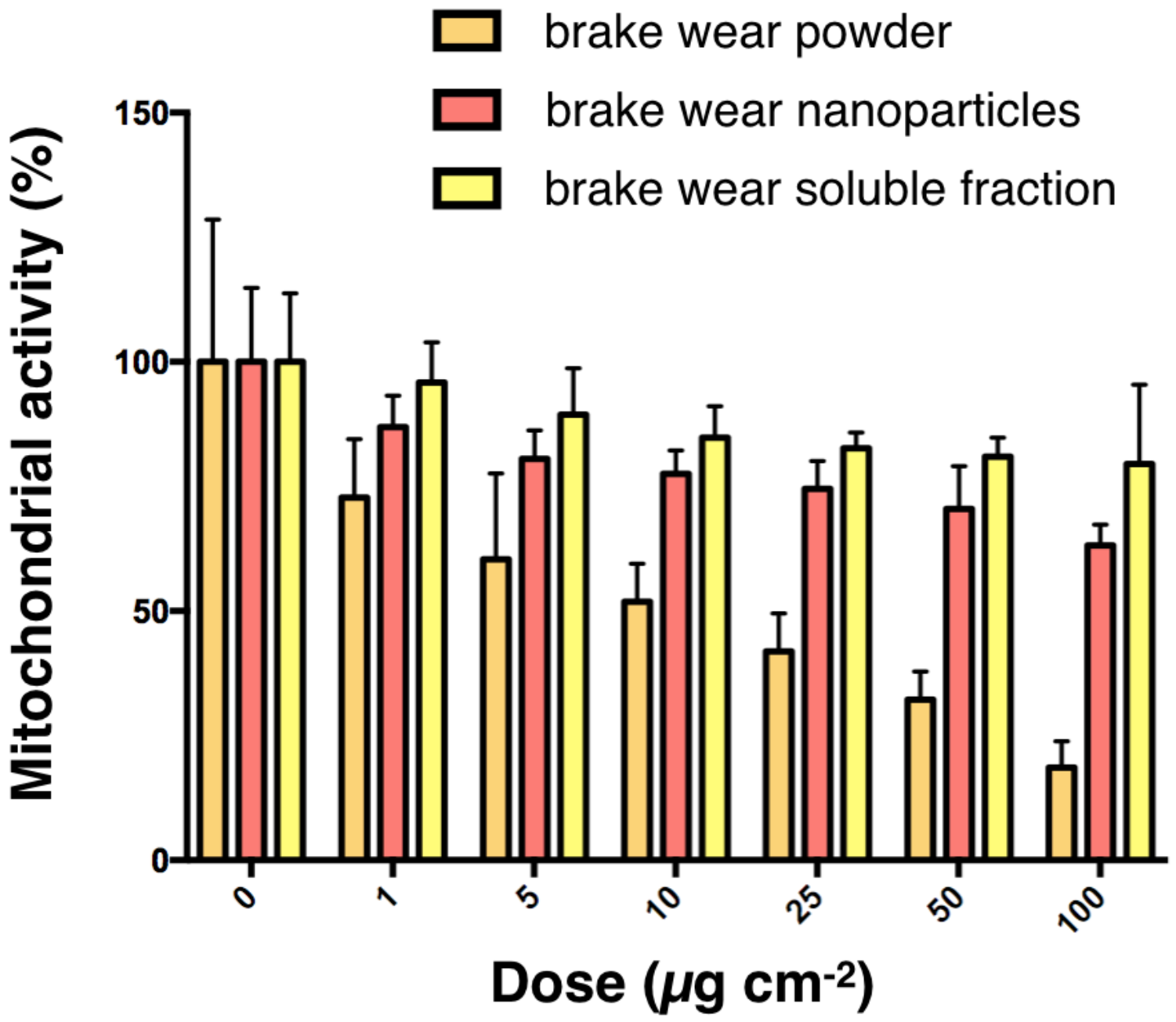

***Figure S9:*** *Cell viability of Calu-3 cells exposed for 24h to different concentrations of BWP, their nanosize fraction and the corresponding soluble fraction. Mitochondrial activity reflecting cell viability was assessed using WST-1 assay. The soluble fraction was the supernatant recovered after BWP sonication, filtration and centrifugation.*

## Supporting Information Figure S10 – Application of the in vitro sedimentation, diffusion and dosimetry model to the brake wear nano- and microparticles

---

Dosimetry simulation for 1.47 µm BWP particles

| Input variables used in the simulation | Unit | Values |
|---|---|---|
| Solvent dynamic viscosity | Pa s | 0.000783 |
| Density of media | g $cm^{-3}$ | 1.0091 |
| Temperature | ° C | 37.0 |
| Density of raw material | g $cm^{-3}$ | 5.02 |
| Diameters ($d_{OM}$) of each existing particle/agglomerate species | µm | 1.47 |
| Fractions of agglomerate species | .. | 1 |
| Agglomerate effective density | g $cm^{-3}$ | 1.4 |
| Height of suspension column | mm | 3 |
| Initial total concentration of material | mg $cm^{-3}$ | 0.1 |
| Total time of simulation | h | 2 |
| Height of subcompartment (simulation element) | mm | 0.001 |

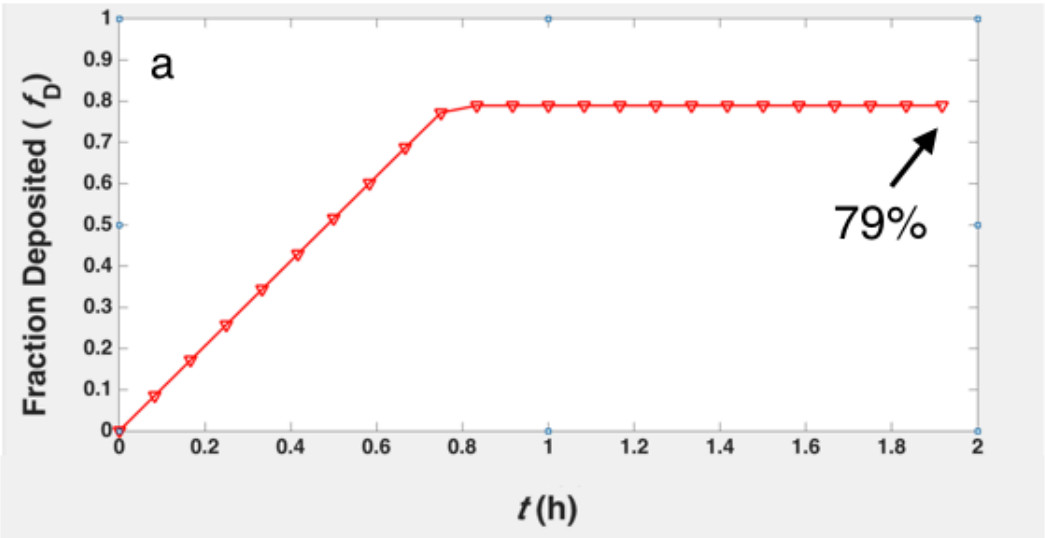

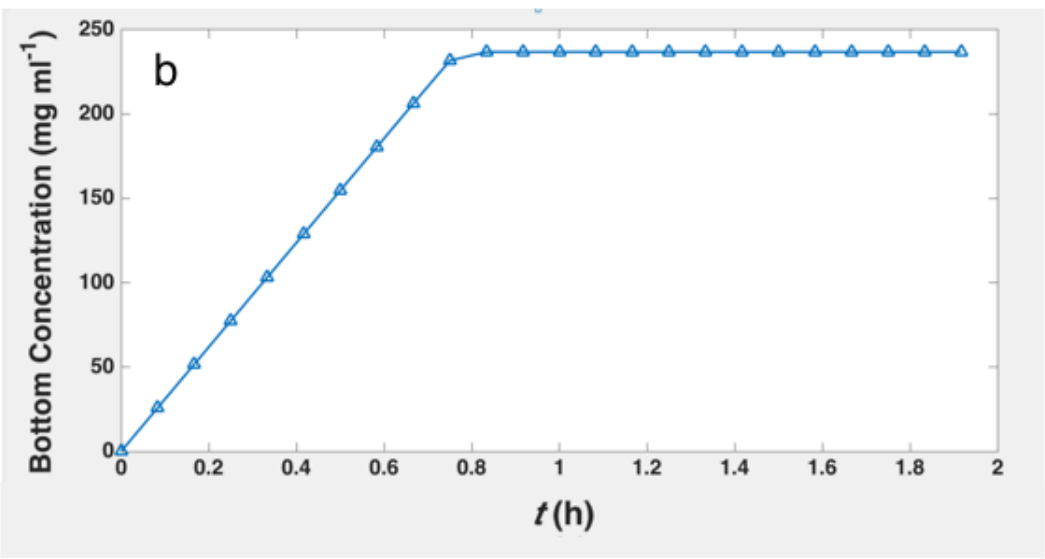

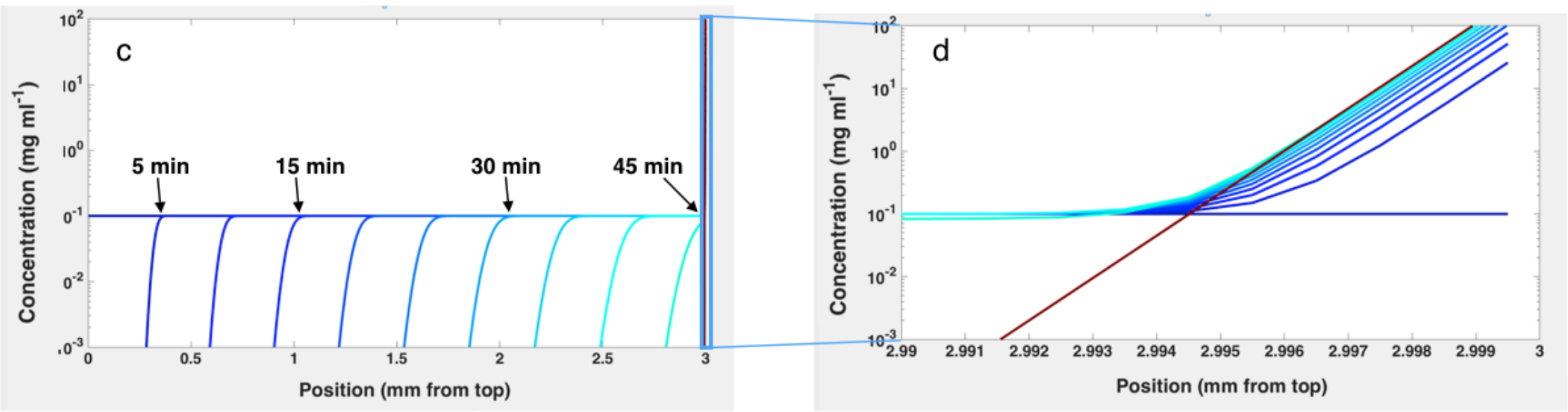

***Figure S10A: Table*** *- List of input parameters injected in the in vitro sedimentation, diffusion and dosimetry model reported by DeLoid and coworkers (DeLoid G.M. et al. Nat. Protoc. 2017, 12, 355-371) for 1.5 µm brake wear particles. The size distribution refers to the data from Fig. S1.* ***a)*** *The fraction of deposited particles as a function of the time. It was found that after 45 min of incubation the dose delivered to the cells saturates at 79% of the total administered dose.* ***b)*** *Bottom concentration as a function of the time. The initial concentration in the dispersion was set at 0.1 g $L^{-1}$ (corresponding to 1 mg $cm^{-3}$, see Table).* ***c and d)*** *Time evolution of the concentration as a function of the position in the sample. The abscissa 3 mm represents the bottom of the cell culture well. The kinetics shows that the sedimentation occurs rapidly, the concentration profiles stabilizing after 45 min. Note that in d) the abscissa has been expended from 2.99 to 3 mm to emphasize that most particles are at the bottom of the well.*

# Dosimetry simulation for the BWP nano-sized fraction

| Input variables used in the simulation | Unit | Values |
|---|---|---|
| Solvent dynamic viscosity | Pa s | 0.000783 |
| Density of media | g cm$^{-3}$ | 1.0091 |
| Temperature | ° C | 37.0 |
| Density of raw material | g cm$^{-3}$ | 5.02 |
| Diameters (dH) of each existing particle/agglomerate species | nm | 170 / 320 / 1000 |
| Fractions of agglomerate species | .. | 0.13 / 0.68 / 0.19 |
| Agglomerate effective density | g cm$^{-3}$ | 1.4 |
| Height of suspension column | mm | 3 |
| Initial total concentration of material | mg cm$^{-3}$ | 0.1 |
| Total time of simulation | h | 24 |
| Height of subcompartment (simulation element) | mm | 0.001 |

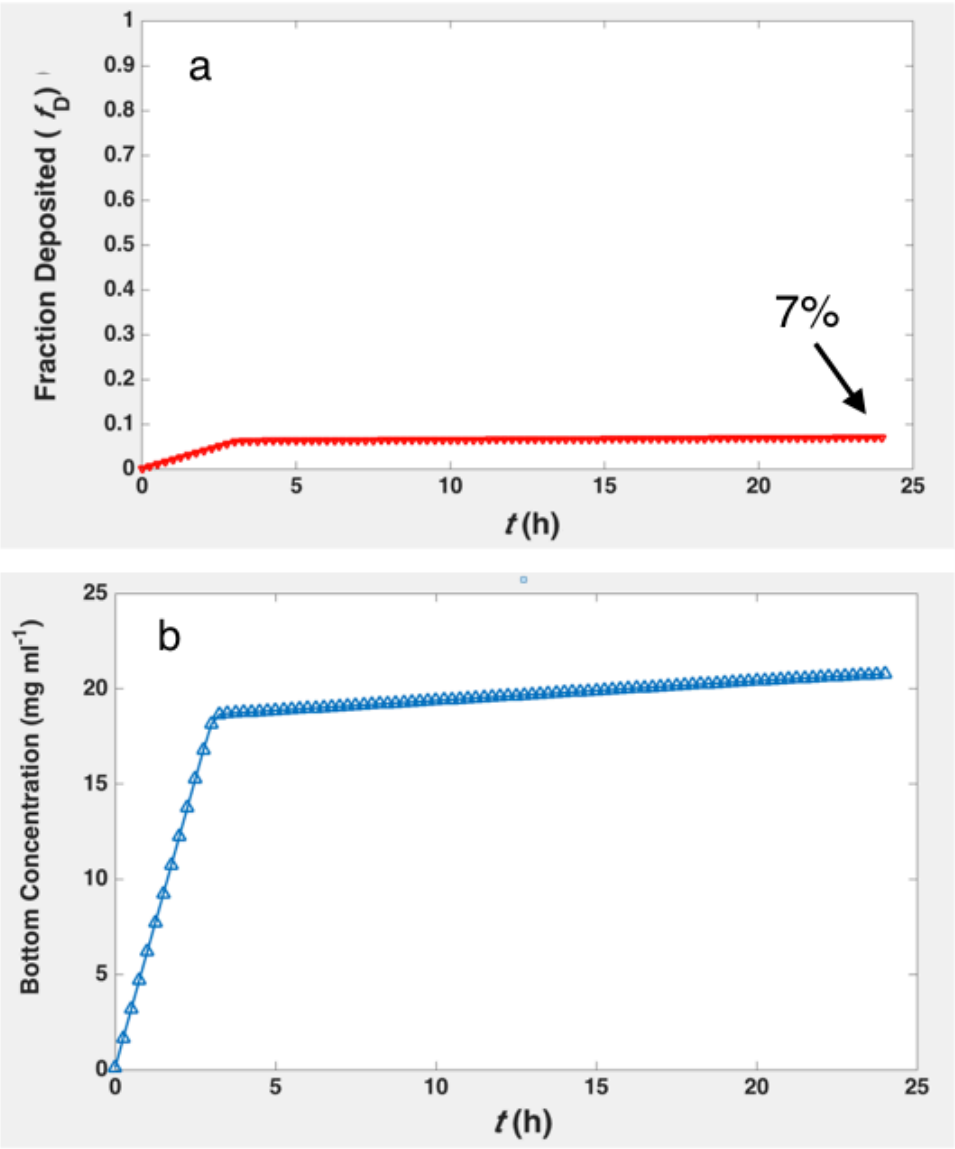

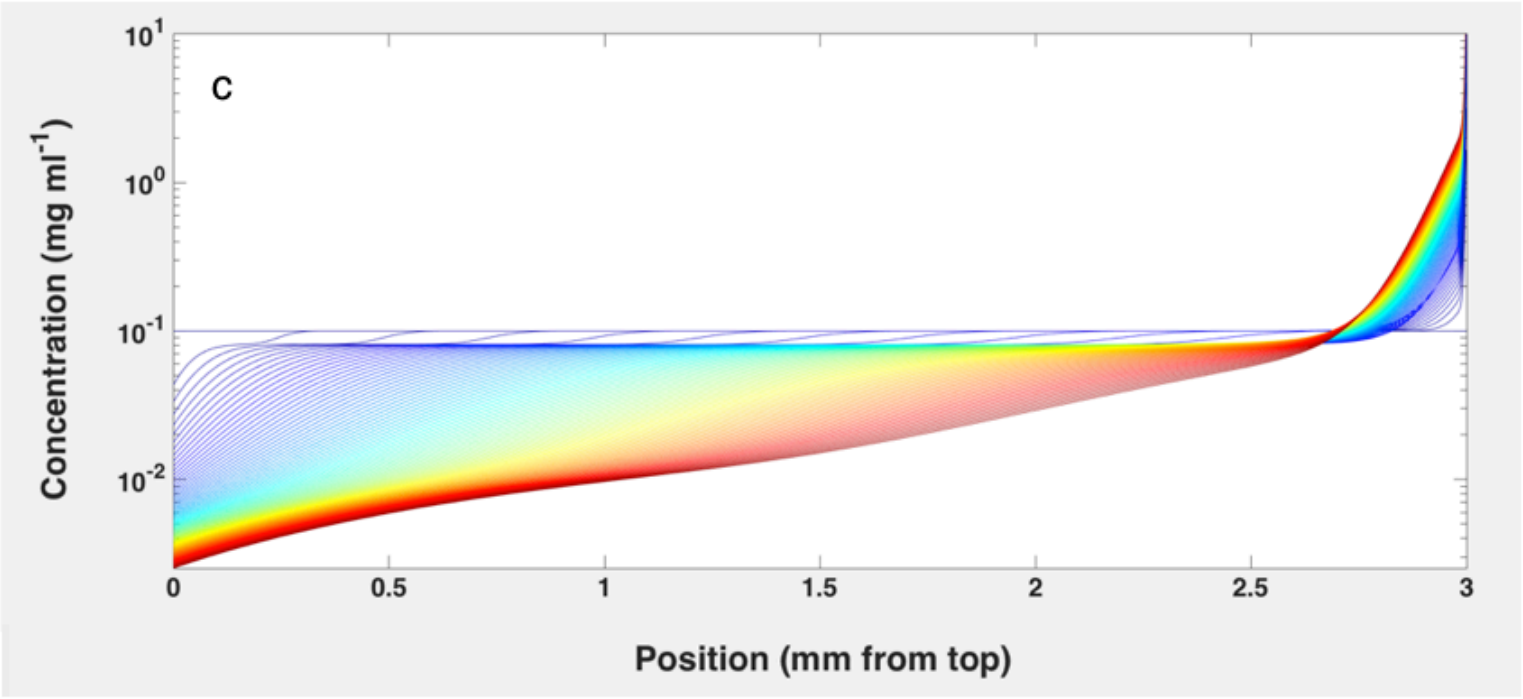

***Figure S10B: Table*** *- List of input parameters injected in the in vitro sedimentation, diffusion and dosimetry model reported by DeLoid and coworkers (DeLoid G.M. et al. Nat. Protoc. 2017, 12, 355-371) for the brake wear nano-sized fraction. The size distribution refers to the light scattering data in Fig. S5c.* ***a)*** *The fraction of deposited particles as a function of the time shows a saturation around 7% of the total administered dose after 24 h.* ***b)*** *Bottom concentration as a function of the time. The initial concentration in the dispersion was set at 0.1 g L*$^{-1}$ *(corresponding to 1 mg cm*$^{-3}$*).* ***c and d)*** *Time evolution of the concentration as a function of the position in the sample. The abscissa 3 mm represents the bottom of the cell culture well. The different colored curves correspond to different snapshots of gradient concentration.*

## Supporting Information Figure S11 – Rescaling toxicity assays with the effective dose

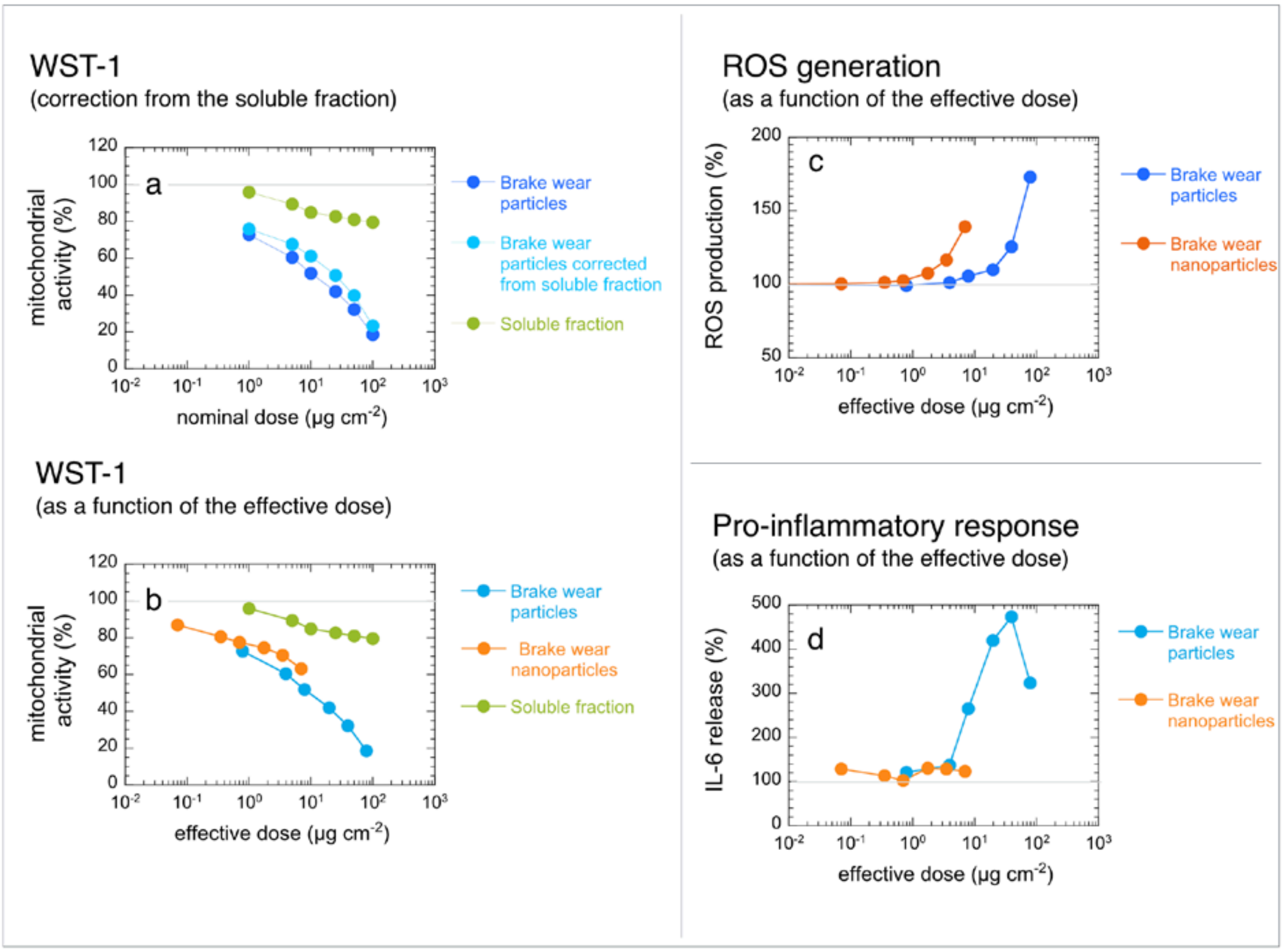

***Figure S11:*** a) Cell viability of BWP particles corrected from the soluble fraction. b) Mitochondrial activity of brake wear nano- and macroparticles expressed as a function of the effective dose. c) Reactive oxygen species production detected by DCFH probe oxidation after 4 h exposure *versus* effective concentration for BWP and its nano-sized fraction. d) Same as in c) for the IL-6 cytokine release measured by ELISA after a 24 h exposure.